\documentclass[useAMS,usenatbib,fleqn,letterpaper]{mn2e_mod} 
\usepackage{graphicx}
\usepackage{subfigure}
\usepackage{epsfig}
\usepackage{natbib}
\usepackage{amsfonts}
\usepackage{mathtools}
\usepackage{amsbsy}
\usepackage{amssymb}
\usepackage{balance}
\usepackage{times}
\usepackage{soul}
\newif\ifAMStwofonts
\AMStwofontstrue
\newcommand{\bhl}{}
\newcommand{\ehl}{}

\newcommand{\beq}{\begin{equation}}
\newcommand{\eeq}{\end{equation}}

\sodef\an{}{.13em}{0.5em plus1em}{2em plus.1em minus.1em}

\newcommand{\cen}{\mbox{Cen~X-4}}
\newcommand{\xmm}{\textit{XMM--Newton}}
\newcommand{\nust}{\mbox{\textit{NuSTAR}}}

\def\rin{$r_{\rm m}$}

\def\rc{$r_{\rm c}$}
\def\del2r{$\Delta r_{\rm 2}$}

\def\fbl{$\mathcal{F}_{\rm bl}$}
\def\fflow{$\mathcal{F}_{\rm flow}$}

\title[Constraining the radiative efficiency of a RIAF]
{The radiative efficiency of a radiatively inefficient accretion flow}

\author[Caroline R.\ D'Angelo et al.]
       {C.\ R.\ D'Angelo,$^{1}$\thanks{\hspace{-0.12cm}E-mail: dangelo@strw.leidenuniv.nl} J.\
         K.\ Fridriksson,$^2$ C.\ Messenger$^3$ and  A.\ Patruno$^{1,4}$\vspace{0.15cm}\\
$^1$Leiden Observatory, Leiden University, Postbus 9513, NL-2300 RA
 Leiden, the Netherlands\\
$^2$Anton Pannekoek Institute for Astronomy, University of Amsterdam,
Postbus 94249, NL-1080 GE Amsterdam, the Netherlands\\
$^3$SUPA, School of Physics and Astronomy, University of Glasgow, Glasgow G12 8QQ, UK\\
$^4$ASTRON, The Netherlands Institute for Radio Astronomy, Postbus
  2, NL-7900 AA Dwingeloo, the Netherlands} 
\date{Accepted 2015 February 27. Received 2015 February 26; in original form 2014 October 14}

\volume{449}
\pagerange{2803--2817}
\pubyear{2015}

\begin{document}

\maketitle

\label{firstpage}

\begin{abstract}
A recent joint \xmm/\textit{Nuclear Spectroscopic Telescope Array }(\textit{NuSTAR}) observation of the accreting neutron star \cen\ ($L_{\rm X}\sim10^{33}{\rm~erg~s}^{-1}$) revealed a hard power-law component ($\Gamma\sim1$--1.5) with a relatively low cut-off energy ($\sim$10~keV), suggesting bremsstrahlung emission. The physical requirements for bremsstrahlung combined with other observed properties of \cen\ suggest the emission comes from a boundary layer rather than the accretion flow. The accretion flow itself is thus undetected (with an upper limit of $L_{\rm flow}\lesssim0.3 L_{\rm X}$). A deep search for coherent pulsations (which would indicate a strong magnetic field) places a 6~per cent upper limit on the fractional amplitude of pulsations, suggesting the flow is not magnetically regulated. Considering the expected energy balance between the accretion flow and the boundary layer for different values of the neutron star parameters (size, magnetic field, and spin) we use the upper limit on $L_{\rm flow}$ to set an upper limit of $\varepsilon\lesssim0.3$ for the intrinsic radiative efficiency of the accretion flow for the most likely model of a fast-spinning, non-magnetic neutron star. The non-detection of the accretion flow provides the first direct evidence that this flow is indeed `radiatively inefficient', i.e. most of the gravitational potential energy lost by the flow before it hits the star is not emitted as radiation.
\end{abstract}


\setcounter{page}{2803}

\section{\hspace{-0.1cm}\an{Introduction}}
It is still not established how accretion flows organize themselves at
the very low luminosities observed in many accreting black hole and
neutron star systems. When such sources are very bright [$L_{\rm
  X}\gtrsim10$ per cent of the Eddington luminosity; $L_{\rm Edd}
\equiv 4\upi GM_* m_{\rm p} c/\sigma_{\rm
  T}=1.8\times10^{38}$~$(M_*/1.4\,\textrm{M}_{\sun}) {\rm~erg~s}^{-1}$] the
accretion flow is an optically thick, geometrically thin accretion
disc, as envisioned by early authors \citep{1972A&A....21....1P,
  1973A&A....24..337S}, and the energy spectrum is dominated by a
blackbody that peaks at hard ultraviolet energies (for supermassive
black holes) or soft X-rays (for stellar mass black holes and neutron
stars). This disc is {\em radiatively efficient}: the energy liberated
by the inspiralling gas is efficiently radiated away. In a thin disc
the luminosity scales linearly with the accretion rate. However, the
luminosity of accreting black holes and neutron stars can vary by more
than eight orders of magnitude. When the luminosity drops below
$10^{-2}$--$10^{-3}$~$L_{\rm Edd}$ the X-ray energy spectrum is
dominated by a power law extending to high energies, typically
$\gtrsim$50--100~keV (e.g. \citealt{2007A&ARv..15....1D}). The
optically thin gas that produces this component is $\sim$1000 times
hotter than what the classic `thin disc' \citep{1973A&A....24..337S}
solution would predict for the observed luminosity, but the geometry,
large-scale dynamic behaviour, and accretion rate of the gas are
uncertain.

More than 30 yr ago, \cite{1982Natur.295...17R} proposed that this
change in spectral behaviour in accreting black holes and neutron
stars could be caused by a radical change in the accretion flow's
structure. They noted that below a critical density the accreting gas
can no longer cool efficiently, so that the accretion flow will
inflate vertically from a cool, dense thin disc into a much hotter,
optically thin torus. As a result, the accretion flow becomes {\it
  radiatively inefficient}, which means that a large fraction of the
gravitational potential energy in the gas is not lost to radiation but
instead converted into some other form of energy (such as kinetic or
internal potential energy). Since accreting black hole and neutron star
accreting systems spend the vast majority of the time in this
low-luminosity state, determining the physical structure and accretion
rate of the flow is necessary to understand much of their
evolution and influence on their environments, e.g. to predict black
hole growth and spin evolution (from accreted matter) as well as
feedback into their surroundings. In neutron stars, the relationship
between luminosity and accretion rate is needed to understand the spin
and mass evolution. Continued accretion at low rates can also produce
emission from the stellar surface, which complicates the
interpretation of observations of quiescent neutron stars as evidence
of deep crustal heating (e.g. \citealt{1998ApJ...504L..95B}).

Unfortunately there is no unique self-consistent dynamic model of a
radiatively inefficient accretion flow (hereafter RIAF). The large
scale height of the flow can result in large-scale radial and vertical
gas motion, which can correspond to global energy and momentum
transport, so that the local temperature and density of the flow
become determined by the global flow properties. The uncertainty over
energy transport relates closely to another major unsolved question:
since the gas must lose a large amount of gravitational potential
energy in order to accrete, where does the `hidden'
(i.e. non-radiated) energy go?

Since the early 1990s there has been considerable theoretical interest
in RIAF solutions, mainly focused on black hole accretion. The most
well known one is probably the `advection-dominated accretion flow'
(ADAF) solution (e.g. \citealt{1994ApJ...428L..13N}), in which the
majority of the accretion energy is advected inwards as protons in the
gas are heated to virial temperatures. Other RIAF solutions propose a
reduced mass accretion rate close to the central object. One popular
set of solutions assumes that most of the energy liberated close to
the central object is advected outward and is used to launch a strong
outflow from the outer parts of the accretion flow (called `ADIOS', or
adiabatic inflow--outflow solution;
\citealt{1999MNRAS.303L...1B,2012MNRAS.420.2912B}). It has also been
suggested that the flow could become `convection' dominated (CDAF;
with large-scale vertical or radial convection cells and a small net
inward accretion rate; e.g. \citealt{2000ApJ...539..809Q}), or
`magnetically' dominated (in which a poloidal magnetic field is
advected inwards from large distances and creates a magnetic barrier
against accretion; \citealt{1974Ap&SS..28...45B,
  2003PASJ...55L..69N}). A considerable amount of accretion energy
must also be used to launch the strong jets and outflows observed in
the low-luminosity state of both black holes and neutron stars
(e.g. \citealt{2004MNRAS.355.1105F, 2005ApJ...635.1203M}).

The best observational constraint on the existence of a RIAF comes
from Sgr A*, the black hole at the centre of the Galaxy. Sgr A* has a
typical non-flaring luminosity of $L_{\rm X}\sim10^{33}$~erg~s$^{-1}$,
or about 10$^{-11}$~$L_{\rm Edd}$ (for a mass of
$4.1\times10^{6}$~M$_{\sun}$; \citealt{2008ApJ...689.1044G}). \bhl This
is far below $L_{\rm X}\sim10^{41}$~erg~s$^{-1}$, the luminosity
corresponding to the inferred accretion rate of
$10^{-5}$~M$_{\sun}$~yr$^{-1}$ at the Bondi radius (estimated from the
rate of gas capture from massive stellar winds in the Galactic Centre;
e.g. \citealt{2000ApJ...539..809Q})\ehl. Both radio observations of
the inner 100~$R_{\rm S}$ \citep{2007ApJ...654L..57M} and X-ray
observations of the inner $10^{5}$~$R_{\rm S}$ around the black hole
\citep{2013Sci...341..981W} show low-density gas consistent with being
produced by a RIAF. ($R_{\rm S} \equiv 2GM/c^2$ is the Schwarzschild
radius corresponding to the event horizon of a non-spinning black hole
of mass $M$). However, the central argument for a RIAF remains the
strong imbalance between the accretion rate at the Bondi radius and
the central luminosity. Without probing intermediate radii (where gas
could be stored, as in transient binaries, or expelled) it is not
possible to directly test the radiative inefficiency of the gas
closest to the black hole.

Neutron stars in accreting binary systems show similar spectral and
variability properties to those of black hole binaries (suggesting
similar accretion physics), but unlike black holes have a hard surface
where matter ultimately settles. The impact of gas with the surface
will release a tremendous amount of energy which will likely be
radiated efficiently away, so that the luminosity from the stellar surface
should be a good proxy for the accretion rate on to the star and hence
in the innermost regions of the accretion flow
(e.g. \citealt{1995ApJ...452..710N}). For example, if the majority of
the accretion energy is advected inwards (ADAF) the surface of the
neutron star will be much brighter than if the energy is used to expel
gas in an outflow (as in e.g. ADIOS;
\citealt{1999MNRAS.303L...1B}). Indeed, the relative brightness of
neutron star transients compared with black holes has been used as
evidence of the existence of both an event horizon
\citep{1997ApJ...478L..79N} and an ADAF \citep{2003MNRAS.342.1041D}.

Using neutron star binaries to constrain radiative efficiency requires
understanding how the energy spectrum is produced. At very low
luminosities in quiescent neutron star binaries, two spectral X-ray
components are typically observed: a thermal (blackbody-like) one and
a power-law one. It is generally accepted that the thermal component
is produced by the neutron star surface, either as a result of heating
by accreting gas or by nuclear processes in the deep crust
\citep{1998ApJ...504L..95B}. The power-law component most likely
originates either in the boundary layer close to the neutron star or
in the accretion flow itself \citep{1999ApJ...520..276M}.\footnote{An
  alternative suggestion -- that the quiescent X-ray emission comes
  from the impact of a pulsar wind from the neutron star with the
  outer accretion disc \citep{2000ApJ...541..849C} -- has been
  disfavoured by more recent observations
  \citep{2013MNRAS.436.2465B,2014ApJ...797...92C}}

If the power-law emission originated from a radiatively inefficient
accretion flow, the neutron star surface (blackbody component) would
be expected to be much brighter (potentially orders of magnitude) than
the emission from the flow. In contrast, the thermal and power-law
components for the faintest neutron star binaries (typically radiating
at $\sim$$10^{-5}$--$10^{-6}$~$L_{\rm Edd}$) are often comparable
(\citealt{1998A&ARv...8..279C};\bhl\ see also fig.\ 5 of
\citealt{2004MNRAS.354..666J})\ehl. Furthermore, since the surface
emission should be radiatively efficient, \cite{1999ApJ...520..276M}
demonstrated that the very low luminosity from quiescent neutron stars
requires a much lower mass transfer rate from the companion star than
is expected from mass transfer models. To resolve this imbalance, they
proposed that the fast-rotating magnetic field of these neutron stars
expels most of the accretion flow (the `propeller' scenario;
\citealt{1975A&A....39..185I}). The net accretion rate on to the star
is thus several orders of magnitude lower than the accretion rate in
the flow. Although they applied this idea to the ADAF model in
particular, a strong propeller (or some other mechanism to expel gas
that is unrelated to the accretion flow) is generically required for
any radiatively inefficient flow if the power-law component is
generated by the accretion flow, otherwise the surface will be much
brighter than the flow. However, many low-luminosity neutron star
binaries show no signs of magnetic activity (such as X-ray
pulsations), and recent work has demonstrated that the magnetic field
required for the `propeller' scenario to work in quiescent neutron
stars is about 10 times as large as those in accreting millisecond
X-ray pulsars, where pulsations are seen \citep{2013MNRAS.436.2465B}.

A recent observation of a nearby accreting neutron star, \cen, has
offered new insights into the accretion properties of low-luminosity
neutron star binaries. In simultaneous \textit{Nuclear Spectroscopic
  Telescope Array} (\nust) and \xmm\ observations of \cen\ in 2013
January, \cite{2014ApJ...797...92C} found evidence of a break or
cut-off in the hard X-rays around 10~keV, while the index of the
power-law component below 10~keV was quite hard:
$\Gamma\sim$~1--1.5.
They considered different physical scenarios for the origin of the
high-energy emission, before focusing on a scenario in which the
spectrum is produced by bremsstrahlung radiation from a hot massive
outflow at a large distance from the star, roughly as predicted by the
model suggested by \cite{1999MNRAS.303L...1B}.

In this paper, we propose that the power-law component seen in \cen\
and similar quiescent neutron star binaries is most plausibly produced
by bremsstrahlung emission from the boundary layer on the surface of
the star. This suggests that {\em all} the observed emission is coming
from near the surface of the star, and the power-law and thermal
components are directly coupled. Our interpretation naturally fits the
observed spectrum, and also explains the energy balance and
covariation of the two components without invoking a magnetic
propeller. Most significantly, if all the observed emission comes from
 a boundary layer, then the flow itself is completely undetected,
i.e. it is indeed radiatively inefficient. This would be the most
direct, model-independent evidence for radiative inefficiency in any
accreting system.

The paper is organized as follows. In Section \ref{sec:models} we
review how much emission is expected to come from a neutron star
boundary layer and accretion flow, and how the neutron star's
properties affect this balance. In Section \ref{sec:CenX4} we
summarize the observed properties of \cen\ and present a partial
reanalysis of the data presented in \cite{2014ApJ...797...92C} in
order to constrain the presence of an additional spectral
component. We also present the results of a search for coherent
pulsations (which would indicate magnetically channelled
accretion). In Section \ref{sec:phys_proc} we reinvestigate the
accretion-flow model proposed by \cite{2014ApJ...797...92C} and
outline its difficulties in explaining the observation. We then
demonstrate that the most plausible model for the power-law component
is bremsstrahlung from the star's surface. We also set limits on the
scale height and density of the boundary layer. (A more complete
analysis of the physical constraints on the boundary layer imposed by
this observation will follow in a separate paper.)  Finally, in
Section \ref{sec:RIAF} we calculate upper limits for the intrinsic
radiative efficiency of the accretion flow, building on the spectral
analysis presented in Section \ref{sec:CenX4}. In Sections
\ref{sec:discussion} and \ref{sec:conclusion} we discuss briefly the
implications of our results, summarize our conclusions, and explore
how future observations of low-level accretion on to neutron stars can
be used to put better constraints on the radiative efficiency of the
accretion flow.

\section{\hspace{-0.1cm}\an{How accretion energy is liberated in neutron stars}}
\label{sec:models}
What fraction of the infalling gas's gravitational potential energy
will be converted to radiation in the boundary layer of the neutron
star? This question is central to the argument of this paper: to put
constraints on the intrinsic radiative efficiency of the accretion
flow (i.e. the amount of gravitational potential energy that the
accretion flow itself radiates away), we need to know how much energy
is still stored in the gas when it reaches the boundary layer. The
division of energy between the boundary layer and accretion flow can
be altered both by the properties of the neutron star (its size, spin,
and magnetic field, as described in Sections \ref{sec:magnetosphere}
and \ref{sec:other}) and by the properties of the boundary layer and
accretion flow.

The total rate of gravitational potential energy released \bhl in a
Newtonian potential\ehl\ by matter accreting at a rate $\dot{M}$
on to a star of mass $M_*$ and radius $R_*$ is
\beq
\label{eq:edot}
\dot{E}_{\rm pot} = \frac{GM_*\dot{M}}{R_*}, 
\eeq which can be divided
into the fraction of energy available to be radiated by the flow,
\fflow, and the boundary layer, \fbl.\footnote{Note that throughout
  this paper we define radiative efficiency $\mathcal{F}$ as the
  fraction of gravitational potential energy released as
  radiation. This is a different quantity from $L/\dot{M} c^2$,
  the fraction of the accretion flow's {\em rest-mass energy} that is
  radiated, which is also sometimes called the `radiative efficiency'
  of an accretion flow (and is $\sim$10 per cent for a black hole).} 

In this paper we define the boundary layer as the radius where the
angular frequency of the gas reaches a maximum, usually very close to
the star, so that $b/R_* \ll 1$, where $b$ is the width of the layer
\citep{1973A&A....24..337S,2002apa..book.....F}. Provided that the
accretion flow remains nearly Keplerian, the kinetic energy of a test
particle at the boundary layer is $\sim$$0.5 m v^2_{\rm K,*}$ (where
$v_{\rm K,*}$ is the Keplerian velocity at the star's surface), which
will mainly be converted to radiation when it hits the star. As long
as the boundary layer is very thin, the two components \fbl\ and
\fflow\ will thus be roughly equal, so that \fbl\ $\simeq$ \fflow\
$\simeq 0.5$ (e.g. \citealt{2002apa..book.....F}). However, if the
boundary layer is {\em not} thin, and $b \gtrsim R_*$, the gas will
also have additional gravitational potential energy \mbox{$GM_*m/(R_*+b)$} at
the top of the boundary layer, which will then change the balance
between \fflow\ and \fbl, increasing \fbl\ by a factor of \mbox{$(R_*+b)/R_*$}
 \citep{2001ApJ...547..355P}. If the boundary layer is very physically extended
the flow will thus contribute much less radiation to the total
spectrum, making it harder to establish that the accretion flow is
intrinsically radiatively inefficient.

\subsection{\hspace{-0.15cm}How thick is the boundary layer?}
At low accretion rates the assumption that $b
\ll
R_*$ may not be valid. In their study of accretion on to white dwarfs,
\cite{1993Natur.362..820N} found that the boundary layer thickness can
increase substantially as $\dot{M}$
decreases, mainly due to a thermal instability that causes the
boundary layer to heat up and expand. However, their results (and the
later study of neutron star boundary layers of
\citealt{2001ApJ...547..355P}) may have considerably overestimated the
boundary layer temperature due to an incomplete treatment of radiative
cooling \citep{2002AstL...28..150G}, thus overestimating the boundary
layer thickness. Observations of neutron star low-mass X-ray binaries
(LMXBs) suggest that the boundary layer becomes optically thin below
$\sim$$0.01$~$L_{\rm
  Edd}$,
as evinced by the lack of a strong blackbody component in the X-ray
spectrum (e.g. \citealt{1999ApL&C..38..121R}), which would also
suggest an extended, optically thin boundary layer. To our knowledge,
no theoretical work has studied the formation of a boundary layer at
the low accretion rates seen in \cen, so it is difficult to make
anything more than basic predictions for boundary layer thickness.

In this paper our aim is to demonstrate that the accretion flow in
\cen\ is radiatively inefficient. We therefore consider limits on what
the boundary layer properties would be assuming the flow is
radiatively efficient, which can then be excluded by
observation. Following the logic for standard (thin) accretion discs,
in the boundary layer the gas must rapidly decelerate from nearly
Keplerian orbits to corotation with the star. As long as the net
infall velocity is not strongly supersonic, the gas pressure gradient
must balance the gravitational force, so that \beq b \simeq
\frac{c^2_{\rm s}}{v^2_{\rm K,*}}R_* \eeq
(where $c_{\rm s}$ is the gas's sound speed; \citealt{2002apa..book.....F}). If the protons and electrons in the gas
are efficiently coupled, then the temperature of the entire gas is
constrained by the \nust\ observation to be $\sim$$18$~keV \citep{2014ApJ...797...92C},
corresponding to $b\sim 10^2$~cm. However, if the electrons and protons
are not efficiently coupled, the proton temperature could be much
larger -- of order the virial temperature. In this case, \beq c_{\rm
  s}^2 \sim \frac{\alpha v^2_{\rm K,*}}{5} \eeq
(e.g. \citealt{1987A&A...185..155K}), so that $b\sim 0.2\alpha R_*$,
where $\alpha < 1$ is the standard viscosity parameter. We thus assume
that, provided that the flow is quasi-Keplerian close to the star, the
boundary layer will remain small compared with the radius of the star,
so that \fbl~$\simeq$~\fflow~$\simeq0.5$.

\subsection{\hspace{-0.15cm}Accretion on to a magnetosphere at low $\dot{M}$}
\label{sec:magnetosphere}
\begin{figure}
\centerline{\includegraphics[width=85mm]{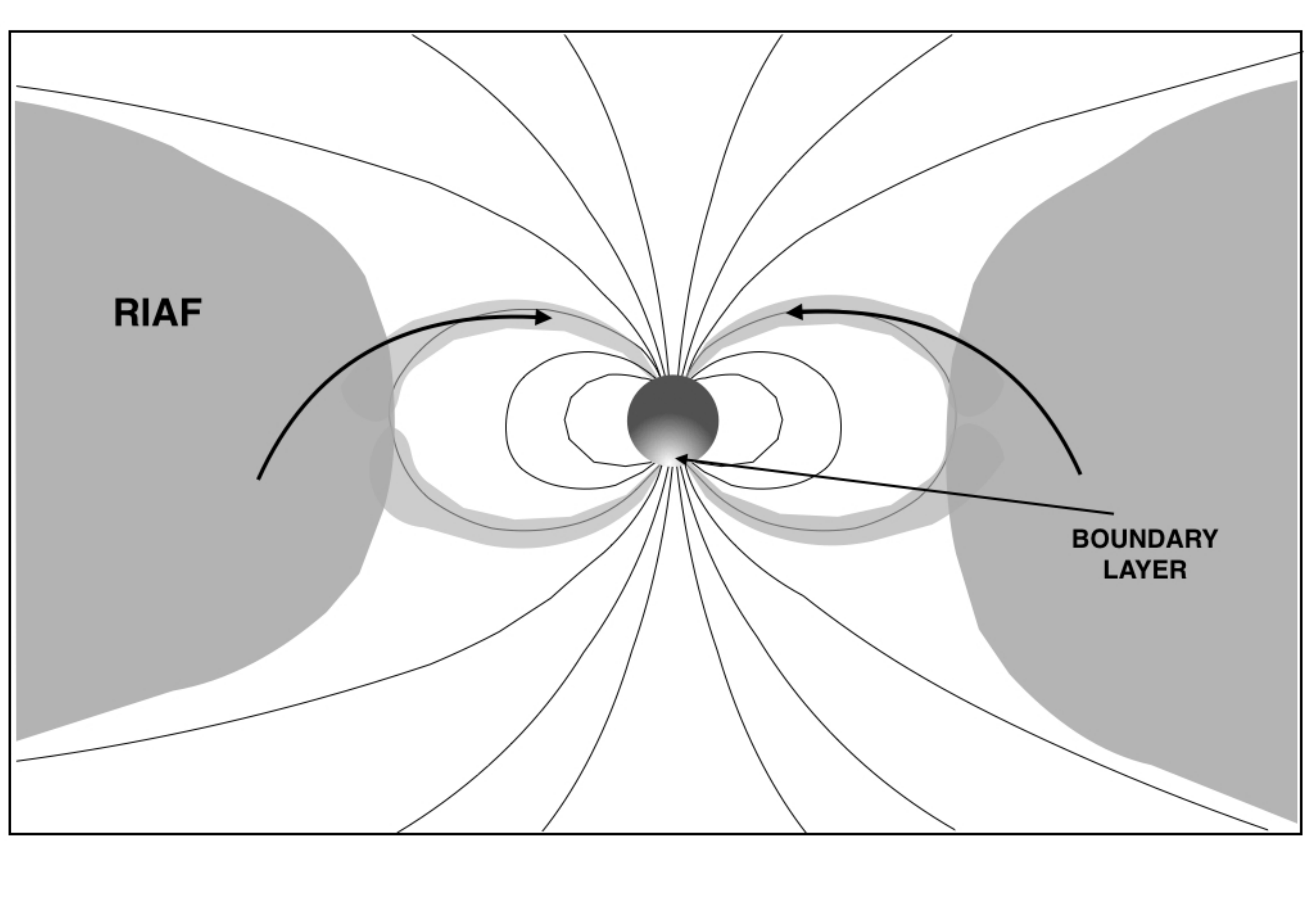}}
  \caption{\label{fig:global} Schematic illustration of accretion from
    a radiatively inefficient flow on to a strong magnetic field. MHD
    simulations suggest that the magnetic field will adopt a mainly
    open configuration, with limited connection between the star and
    the flow. Since the flow does not radiate efficiently while the
    boundary layer does, emission from the flow is not detected, while
    accretion on to the boundary layer produces the observed X-ray
    spectrum in quiescent neutron stars.}
\end{figure}

The largest potential influence on the boundary layer/accretion flow
energy balance will come from a strong stellar magnetic field. A
sketch of a RIAF accreting on to a star with a strong magnetic field
is shown in Fig.~\ref{fig:global}. Accretion from a RIAF on to a
magnetosphere has not been theoretically investigated, but the basic
physics should be similar to magnetospheric accretion from a disc. The
magnetic field will truncate the inner part of the accretion flow at
the `magnetospheric radius' $r_{\rm m}$ and channel accreting
material on to the pole of the star
(e.g. \citealt{1972A&A....21....1P,1977ApJ...217..578G}). The inner
magnetosphere will be nearly evacuated of gas, while the stellar
surface will be significantly brighter than for a non-magnetic
star. The interaction between the disc and the field will also allow an
exchange of angular momentum between the star and the flow.

The magnetic field has two effects on the energy balance between the
flow and the boundary layer. First, since the inner edge of the flow
($r_{\rm m}$) is now much further from the star, \fflow\ will be reduced
by $R_{*}/r_{\rm m}$. Second, if the star is spinning much faster than
the inner edge of the disc, the relative spin rate of the magnetic
field and inner accretion flow can launch an outflow of gas from the
disc (the `propeller' effect; \citealt{1975A&A....39..185I}). This
will dramatically reduce accretion on to the neutron star's surface,
and thus \fbl\ and the luminosity of the boundary layer. 

In an accretion flow surrounding a strong stellar magnetic field, the
inner edge can be defined as the point where the magnetic field is
strong enough to enforce corotation of gas
\citep{1993ApJ...402..593S}:
\begin{align}
r_{\rm m} \mspace{1.5mu} &= \mspace{1.5mu} \left(\frac{\eta\mu^{2}}{4\Omega_*\dot{M}}\right)^{1/5} \nonumber\\
                \mspace{1.5mu} &= \mspace{1.5mu} 2.4\times10^6\,\eta^{1/5} \left(\frac{B_*}{10^8\,\textrm{G}}\right)^{2/5}\left(\frac{R_*}{10^6\,\textrm{cm}}\right)^{6/5} \nonumber\\
               & \mspace{20mu} \times\left(\frac{P_*}{2\times10^{-3}\,\textrm{s}}\right)^{1/5}\left(\frac{\dot{M}}{10^{16}\,\textrm{g}\,\textrm{s}^{-1}}\right)^{-1/5} {~\rm cm}.
\label{eq:rm}
\end{align}
In this equation, $\mu = B_*R_*^3$ is the magnetic moment of the star,
and $\eta \leq 1$ is a dimensionless parameter characterizing the
strength of the toroidal magnetic field induced by the relative
rotation between the disc and dipolar magnetic field. $\Omega_*$
($P_*$) is the star's angular spin frequency (period) and $\dot{M}$ the mass
accretion rate through the disc. The magnetospheric radius is scaled
for a rapidly spinning, weakly magnetized neutron star in outburst (at
$\sim$$0.01\dot{M}_{\rm Edd}$, or $L_{\rm X} \sim 1.8\times10^{36}{\rm
  ~erg~s}^{-1}$). Throughout this paper, we assume a neutron star of
$1.4$~M$_{{\sun}}$ and a radius of $10^{6}$~cm.\footnote{Equation (\ref{eq:rm}) predicts a somewhat
  smaller truncation radius than the more commonly used prescription
  \citep{1972A&A....21....1P,1977ApJ...217..578G} which equates the
  ram pressure of the infalling gas and magnetic pressure of the
  magnetic field of the star without considering the relative rotation
  between the two components. An uncertainty of at least $\sim$2--3
  for \rin\ should be assumed.}

\subsubsection{\hspace{-0.15cm}The magnetic `propeller' model}
\label{sec:propeller}
The biggest uncertainty in establishing \fbl\ and \fflow\ is
determining what happens if the magnetic field of the star spins
faster than the inner edge of the accretion flow. This can happen when
the accretion rate decreases enough that \rin\ (given by equation
\ref{eq:rm}) lies outside the corotation radius, $r_{\rm c} \equiv
(GM_*/\Omega_*^2)^{1/3}$, where the Keplerian disc and star rotate at
the same rate. In this case, the rotating magnetic field acts as a
centrifugal barrier, preventing matter from accreting on to the
star. It is often assumed that the majority of matter accreting
through the disc will be expelled when it hits this centrifugal
barrier (the `propeller' mechanism; \citealt{1975A&A....39..185I}),
thus dramatically reducing \fbl.

A considerable uncertainty in the propeller model is how much gas (if
any) is actually accreted on to the star. Some authors have assumed
that once $r_{\rm
  m} > r_{\rm
  c}$ {\em all} the gas is expelled in an outflow, so that only the
accretion flow itself is visible
(e.g. \citealt{1994ApJ...423L..47S}). In contrast, magnetohydrodynamics
(MHD) simulations of the strong propeller regime tend to show variable
accretion episodes, where a considerable amount of mass is
periodically accreted on to the star
\citep{2004ApJ...616L.151R,2006ApJ...646..304U,2014MNRAS.441...86L}.
In simulations the amount of gas accreted on to the surface decreases
as the propeller gets `stronger' (i.e. the accretion rate decreases
and \rin\ moves further from \rc).  In their interpretation of the
\cen\ quiescent spectrum \cite{2001ApJ...557..304M} also assumed some
gas accretes on to the star. Since \cen's large, variable blackbody
component (interpreted as emission from the stellar surface) is most
likely produced by accretion, the propeller solution can only apply if
some gas accretes on to the star. In our investigation of how the
radiative balance between the star and the flow is affected by a
strong propeller, we use the results of simulations by
\cite{2006ApJ...646..304U} and \cite{2014MNRAS.441...86L}.

\subsubsection{\hspace{-0.15cm}The trapped disc}
\label{sec:trapped}
Recent work investigating the interaction between a magnetosphere and
an accretion disc has suggested that a strong propellered outflow may
not easily form when $r_{\rm m} > r_{\rm c}$
\citep{1993ApJ...402..593S,2010MNRAS.406.1208D}. This is because in
order to efficiently drive an outflow \rin\ must lie significantly
outside \rc, otherwise the differential rotation between the disc and
star is not large enough to accelerate the majority of the infalling
gas to its escape velocity \citep{1993ApJ...402..593S}. Instead, the
disc--field interaction can keep matter confined in the disc, and the
angular momentum added by the disc--field interaction at the inner
edge of the disc is carried outward via turbulent viscosity. As a
result, some matter can continue to accrete on to the star even when
the accretion rate is very low, and a strong propeller outflow does
not form. \cite{2010MNRAS.406.1208D} named such discs `trapped discs',
since the inner edge of the disc remains trapped near \rc\ by the
interaction between the disc and the magnetosphere. This can lead to
episodic bursts of accretion
\citep{1993ApJ...402..593S,2010MNRAS.406.1208D,2012MNRAS.420..416D} as
well as efficient spin-down of the star \citep{2011MNRAS.416..893D}.

\begin{figure}
  \centerline{\includegraphics[width=89mm]{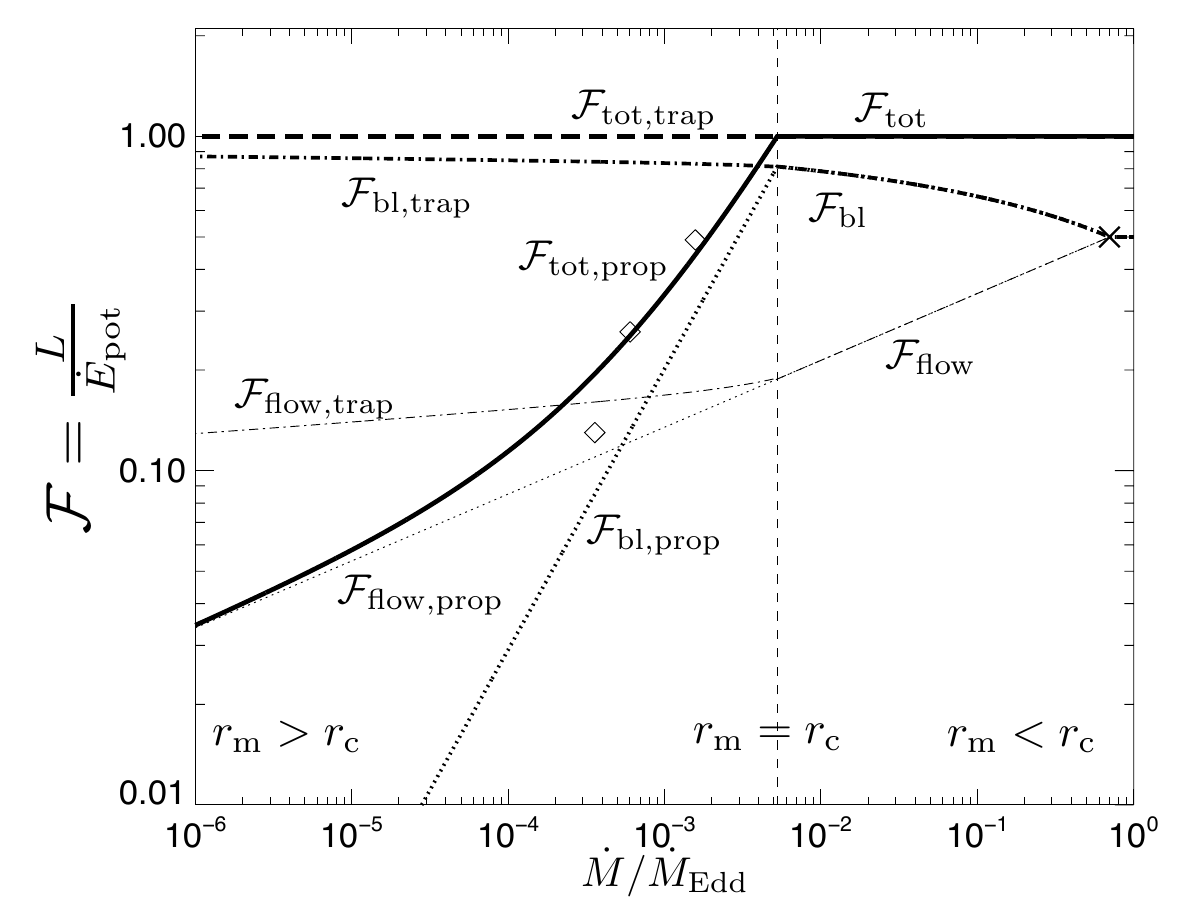}}
  \caption{Radiative efficiency ($\mathcal{F}$) of a 2-ms pulsar with
    a field of $10^8$~G for both the propeller and trapped disc
    models. The radiative efficiency of the boundary layer (bl)
    and flow (flow) is shown as a function of accretion
    rate. At high accretion rates (\rin\ $<$ \rc), the two models are
    equivalent. Above $\dot{M} \simeq 0.7 \dot{M}_{\rm Edd}$ the two
    radiative components are equal since the flow reaches the star
    (disregarding the effects from Sections \ref{sec:size} and
    \ref{sec:spin}). Below $\simeq$$0.7 \dot{M}_{\rm Edd}$ the
    magnetosphere is strong enough to truncate the disc, increasing
    $\mathcal{F}_{\rm bl}$ and decreasing $\mathcal{F}_{\rm flow}$. At
    low accretion rates ($\dot{M} \lesssim
    5\times10^{-3}$~$\dot{M}_{\rm Edd}$), the radiative efficiency is
    very different for the trapped disc [thin (flow) and
    thick (boundary layer) dot--dashed curves; thick
    dashed curve (total); D'Angelo \& Spruit 2012] and when a
    propeller forms [thin (flow) and thick (boundary
    layer) dotted curve; thick solid curve (total)]. The
    propeller model used to estimate \fbl $_{\rm, prop}$ and
    \fflow$_{\rm, prop}$ is from Ustyugova et al. (2006), and the
    diamond points show \fbl $_{\rm ,prop}$ from Lii et al. (2014). In
    a trapped disc, the inner edge of the disc stays close to the
    corotation radius as $\dot{M}$ in the disc decreases, so that gas
    can always be accreted on to the star. In contrast, when $r_{\rm
      m} > r_{\rm c}$ in the propeller scenario, most gas in the disc
    is expelled rather than accreted, so the radiative efficiency
    decreases strongly as a function of $\dot{M}$.}
\label{fig:fbl}
\end{figure}

\begin{figure}
  \centerline{\includegraphics[width=89mm]{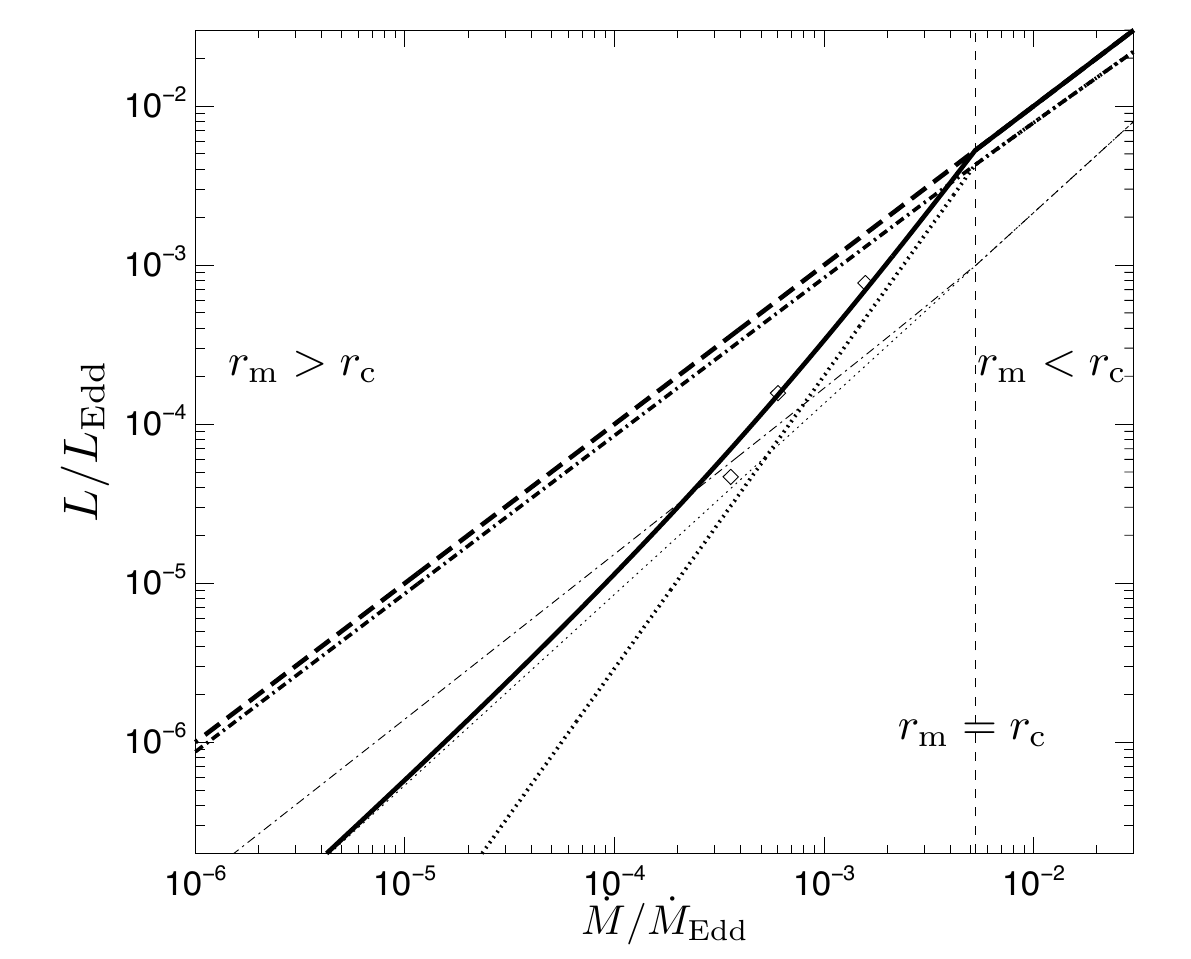}}
  \caption{Luminosity as a function of accretion rate (assuming a
    radiatively efficient accretion flow) for different spectral
    components in a trapped disc and propeller scenario. The
    curve coding is the same as in Fig.~\ref{fig:fbl}. As the
    accretion rate decreases, the luminosity of a trapped disc is
    dominated by the surface component (since the accretion flow is
    truncated), while in a propeller it is dominated by the accretion
    flow (since very little gas is accreted on to the star).}
\label{fig:luminosity}
\end{figure}

The angular momentum from the star added at \rin\ alters the radial
density distribution in the disc. If there is no accretion at all
through the disc the density distribution approaches a `dead disc',
so-called by \cite{1977PAZh....3..262S}, who first derived it. In a
`dead disc' density distribution, the torque added by the magnetic
field at $r_{\rm m}$ matches the rate at which angular momentum is
carried outward by viscous processes through the disc, so that the net
accretion rate on to the star is zero. 

The trapped disc density solution \bhl is the sum of an accreting
component and a `dead disc' component. As the net accretion rate
through the disc decreases \rin\ remains `trapped' very close to \rc\
and the density profile approaches a dead disc one. \ehl The `trapped
disc' is therefore the opposite extreme of the propeller: the angular
momentum from the star is added to the disc rather than expelled in an
outflow, and most of the gas in the disc is ultimately accreted on to
the star. Trapped discs have not yet been properly explored with
numerical simulations, although some evidence of disc trapping was
observed in the `weak propeller' MHD simulations reported by
\cite{2006ApJ...646..304U}.
 
The propeller and trapped disc scenarios result in very different predictions
for how \fbl\ and \fflow\ change with accretion rate. If there is a
strong propeller, only a small amount of gas reaches the stellar
surface, so that \fbl\ is much smaller than \fflow. In a trapped disc
the opposite is true: \fflow\ is much smaller than \fbl\ because the
flow is truncated by the field at a considerable distance from the star.

Fig. \ref{fig:fbl} shows the evolution of \fbl, \fflow, and
\mbox{$\mathcal{F}_{\rm tot}=\mathcal{F}_{\rm bl}+\mathcal{F}_{\rm flow}$}
as a function of accretion rate for a propeller and trapped disc. In
this figure we use a weakly magnetized ($B=10^8$~G) neutron star with
a high spin rate ($P=2$~ms) as our reference model, so that $r_{\rm m}
\sim r_{\rm c} \sim 30$\,km for $\dot{M}/\dot{M}_{\rm Edd}\sim
4\times10^{-3}$. We use the trapped disc model presented by
\cite{2011MNRAS.416..893D} (with $\Delta R/R = \Delta R_{2}/R = 0.1$
and $\eta = 1$; see paper for details of the numerical
parameters). For the propeller regime we use the MHD simulations
reported by \cite{2006ApJ...646..304U}, who empirically fit the
results of simulations of propellering discs of different mean
accretion rate to derive a relationship between the mean $\dot{M}$
through the disc and the net $\dot{M}$ on to the star (and hence
\fbl). The diamond symbols show \fbl\ as found by more recent
simulations by \cite{2014MNRAS.441...86L}. The discrepancy between the
simulations stresses the large uncertainty in how much mass is
actually accreted on to the star in the propeller scenario.

Beginning with the highest accretion rates ($>$$0.7$~$\dot{M}_{\rm Edd}$),
the accretion flow overwhelms the magnetic field and gas accretes
directly on to the star. When $\dot{M}\sim0.7\,\dot{M}_{\rm Edd}$ (the
`$\times$' in Fig.~\ref{fig:fbl}), the magnetic field is strong enough
to disrupt the flow so that \fbl\ increases and \fflow\ decreases as
$\dot{M}$ decreases, following from equation (\ref{eq:rm}). As the
accretion rate drops further and $r_{\rm m} = r_{\rm c}$, the two
possible solutions (trapped disc or propeller) begin to diverge.

In the `propeller' scenario (shown in Fig.~\ref{fig:fbl} as thick
[\fbl] and thin [\fflow] dotted curves),
\fflow~$\propto\dot{M}^{1/5}$, since the inner disc edge moves outward
as $\dot{M}$ decreases. The surface component, \fbl, will decline
steeply with decreasing $\dot{M}$, since the propeller becomes
stronger with decreasing $\dot{M}$ and less gas hits the stellar
surface. \fflow\ will decrease by a factor of 14 if the accretion rate
decreases from the Eddington rate by a factor of $10^6$, and \fbl\ by
more than 800 times so the radiative
contribution from the stellar surface will be completely negligible if
the accretion flow is intrinsically radiatively efficient. The total
available energy for radiating \fflow~$+$~\fbl\ (the thick solid line)
thus decreases rapidly with decreasing $\dot{M}$.  Note that in a
propeller, if $\dot{M}$ decreases from $\dot{M}_{\rm Edd}$ to $10^{-6}\,\dot{M}_{\rm Edd}$, $r_{\rm m}$ will increase from $\sim$$r_*$ to
$\sim$$5\,r_{\rm c}$, meaning that the star would be spinning 10 times
as fast as the inner edge of the disc.

In contrast to the propeller, in the trapped disc at low
$\dot{M}$, \fbl\ and \fflow\ remain approximately constant as
$\dot{M}$ decreases (shown, respectively, by the thick and thin
dot--dashed curves in Fig.~\ref{fig:fbl}), and \mbox{\fbl~$+$~\fflow~$=$~1}
(thick dashed curve) for all accretion rates (since all gas ends up on
the stellar surface). The ratio between \fbl\ and \fflow\ stays roughly
constant because the disc--field interaction `traps' the disc near
\rc, while gas continues to penetrate through the centrifugal barrier
and accrete on to the star.

Fig.~\ref{fig:luminosity} shows the luminosity of the flow and the
boundary layer in both a propelling disc and trapped disc as function
of $\dot{M}$. This assumes both the surface and the flow radiate
efficiently, so the only difference comes from the amount of energy in
each component, \fbl\ and \fflow. The curves are the same as in Fig.
\ref{fig:fbl}. In a trapped disc, the luminosity is dominated by the
surface emission, since the inner disc is truncated by the
magnetosphere. In contrast, most of the emission in a propelling disc
comes from the accretion flow, since the majority of the gas is
expelled from the disc before it can accrete on to the star.

Do \cen\ and other quiescent neutron stars have a strong magnetic
propeller? The strong, fluctuating quasi-blackbody X-ray component most
likely arises from the matter hitting the surface of the star, so the
strength of this component can be used to estimate how much matter
 reaches the surface. In \cen, the power-law X-ray component is nearly as
bright as the thermal component (and they are of similar orders of
magnitude in several other quiescent sources;
\citealt{1998A&ARv...8..279C}). If the power-law component is produced
by radiatively inefficient gas in the accretion flow, there must
therefore be a strong propeller effect, since otherwise the blackbody
component would be much brighter than the accretion flow. In this
case, the near balance between thermal and power-law emission is a
complete coincidence: the propeller expels just enough gas (in \cen\
and other sources) so that the radiatively efficient accretion on to
the surface of the star is as bright as the radiatively inefficient
accretion flow. This is true for any RIAF model. Since the relative
contribution of each spectral component in \cen\ stays roughly
constant with luminosity (while the luminosity change by a factor of 10), the
`propeller' efficiency would also have to scale with accretion rate in
the same way as the RIAF luminosity as the accretion rate changes by
an order of magnitude.

If instead a trapped disc forms, there is no outflow and \fbl\ and
\fflow\ are approximately independent of $\dot{M}$, so that the
accretion rate on to the star matches that in the flow. In this case
it is unlikely that the power-law component originates in the flow,
since Fig.~\ref{fig:fbl} shows that much less accretion energy is
available for radiation than in the stellar surface, whereas \cen\
shows roughly equal contributions from both components.

The power-law component could originate in the boundary layer. In this
case, a strong propeller effect would require a more radiatively
inefficient accretion flow, since the accretion rate in the flow would
be much higher than the amount of gas that actually ends up on the
star. By assuming there is no propeller, the non-detection of the
accretion flow thus sets a robust {\em upper limit} on the intrinsic
radiative efficiency of the flow.

\subsection{\hspace{-0.15cm}Other effects on radiative efficiency}
\label{sec:other}

\subsubsection{\hspace{-0.15cm}Neutron star size}
\label{sec:size}
In the absence of a neutron star magnetic field, the accretion flow
will extend close to the neutron star's surface and general relativity
is needed to determine the energy balance between the accretion flow
and boundary layer. This has been discussed in some detail first by
\cite{1986SvAL...12..117S} and \cite{1988AdSpR...8..135S} (who dealt
chiefly with non-spinning neutron stars), and more recently in
\cite{1998AstL...24..774S}, who consider a rapidly spinning neutron
star.

The neutron star radius can be smaller than the innermost stable
circular orbit -- the smallest radius for which bound orbits exist in
a (non-spinning) Schwarzschild metric, $R_* < R_{\rm isco} = 3R_{\rm
  S}$. Inside this orbit the gas will be decoupled from the rest of
the accretion flow and will fall chaotically on to the surface of the
star. As a result, the boundary layer will be brighter and the
accretion flow dimmer than if the disc extended to the stellar
surface.  \cite{1986SvAL...12..117S} calculated the energy released in
the flow and the boundary layer in a Schwarzschild metric. Expressed as
first-order corrections to the energy balance in Newtonian gravity,
the fraction of energy from the boundary layer and flow will be
\begin{eqnarray}
\mathcal{F}_{\rm bl} &\simeq& 0.5 \left(1 + \frac{R_{\rm S}}{R_{*}}\right),\nonumber \\
\mathcal{F}_{\rm flow} &\simeq& 0.5 \left(1 - \frac{R_{\rm
      S}}{2R_*}\right).
\end{eqnarray}
The uncertainty in the neutron star internal equation of state
introduces uncertainty in $\mathcal{F}_{\rm bl}/\mathcal{F}_{\rm
  flow}$, but even over the entire range of viable equations of state,
this effect will be small except for very soft equations of
state. From consulting e.g. fig.\ 2 of \cite{2001ApJ...550..426L}, a
2~${\rm M}_{\sun}$ star (where the effect is largest) will have
$\mathcal{F}_{\rm bl}/\mathcal{F}_{\rm flow} =$~ 1.8--2.1, while for a
$1.5{~\rm M_{\sun}}$ star $\mathcal{F}_{\rm bl}/\mathcal{F}_{\rm flow}
< 1.8 $. If the star is spinning very rapidly, the Schwarzschild metric
is no longer applicable and $R_{\rm isco}$ will be smaller, further
reducing the imbalance between \fbl\ and \fflow. Even for a massive
non-spinning neutron star and very soft equation of state this effect
is thus small.

\subsubsection{\hspace{-0.15cm}Spinning near break-up}
\label{sec:spin}
If the inner edge of the accretion flow reaches the stellar surface
and the star spins near break-up, angular momentum exchange between
star and disc will also influence the relative flux contributions from
the flow and the stellar surface
\citep{1991ApJ...370..597P,1991ApJ...370..604P}. As the star
approaches its break-up frequency ($\sim$$1230$~Hz for a
1.4~${\rm M_{\sun}}$, 10 km neutron star;
\citealt{2001ApJ...554..322C}), the inner boundary of the disc must
transport stellar angular momentum outward from the star in order for
accretion to continue (known as a `maximally torqued' inner boundary
condition). \cite{1991ApJ...370..604P} and \cite{1991ApJ...370..597P}
demonstrated that this increases the fraction of energy released in
the flow to \fflow~$\simeq$~0.75. 

As gas hits the surface of the star, some of its kinetic energy will
be used to increase the star's rotation rate
(e.g. \citealt{2002apa..book.....F}). As the star's spin increases
this energy becomes considerable, and decreases the fraction of energy
released as radiation by the boundary layer: 
\beq 
\mathcal{F}_{\rm bl} =
\frac{1}{2}\left(1-\frac{\Omega_*}{\Omega_{\rm K,*}}\right)^2, 
\eeq
where $\Omega_{\rm K,*}$ is the Keplerian frequency at the stellar
surface. Although in principle rapid rotation could considerably
reduce radiation from the boundary layer, the fastest confirmed
millisecond pulsar has a spin frequency of 716~Hz
\citep{2006Sci...311.1901H}, which will reduce the boundary layer
luminosity by about a factor of 2 compared with a non-spinning star. The
spin frequency of a neutron star might also be limited by
gravitational wave emission \citep{1998ApJ...501L..89B} or regulation
by the magnetic field \citep{2012ApJ...746....9P} above
$\sim$$1000$~Hz, the effect of spin rate will likely not influence the
surface/disc luminosity balance by more than a factor of $\sim$2--3. In
\cen, the history of strong accretion outbursts (which would increase
the stellar spin rate) and non-detection of a magnetic field (which if
present could spin down the star), both suggest a fast spin period,
which would lead to reduced energy release from the boundary layer. If
only the boundary layer is detected in the X-ray spectrum (and there
is no substantial magnetic field), a rapid spin implies that the
accretion flow is even more radiatively inefficient than if \cen\ is a
slow rotator.

Without any constraints on the size, spin, or magnetic field for the
neutron star, there is thus considerable uncertainty in the possible
energy balance between the accretion flow and the boundary layer,
which will add uncertainty to any attempt to constrain the intrinsic
radiative efficiency of the accretion flow. However, as we outline in
Section \ref{sec:RIAF}, even within these uncertainties, a reasonably
small measured upper limit on $L_{\rm flow}$ can still put some
constraints on the intrinsic radiative efficiency of the flow.

\section{\hspace{-0.1cm}\an{Observations of Cen X-4}}
\label{sec:CenX4}
\cen\ is the brightest quiescent neutron star LMXB,
and was last observed in outburst in 1979
\citep{kaluzienski1980}. Based on the observed X-ray bursts
\citet{1989A&A...210..114C} derived a distance upper limit of
1.2 $\pm$ 0.3 kpc (see also \citealt{gonzales2005} and
\citealt{kuulkers2009}). The orbital period of the binary is 15.1 h
\citep{1989A&A...210..114C}, and the companion is thought to be of
spectral type K3--K7 V \citep{torres2002,davanzo2005}. The source has
a typical X-ray luminosity of ${\sim}10^{32}\textrm{ erg s}^{-1}$, but
is variable on a wide range of time-scales. It shows correlated X-ray,
UV, and optical variability on time-scales down to at least $\sim$100
s, and a change of more than factor of 20 in 0.5--10~keV luminosity has
been observed over a few days \citep{2013MNRAS.436.2465B}. As
mentioned above, the 0.5--10 keV spectrum shows two components -- a
soft blackbody-like component and a harder component well fitted by a
power law with photon index \mbox{$\Gamma \sim$ 1--2} -- which vary together and contribute
roughly equal amounts of flux
\citep{2010ApJ...720.1325C,cackett2013,2013MNRAS.436.2465B}. The
relative proximity of the source means that a high-quality X-ray
spectrum can be obtained even in quiescence, providing a rare
opportunity to study accretion at very low luminosities in detail.

We have reanalysed the simultaneous \xmm\ and \nust\ observations of
\cen\ performed in 2013 and reported in \citet{2014ApJ...797...92C},
in order to put upper limits on the strength of the undetected
accretion flow. In Section~\ref{sec:spectral} we describe the basic
data extraction and spectral fitting we performed; in
Section~\ref{sec:RIAF} we then use these results to test for the
possible presence of an additional spectral component from the
accretion flow. The analysis procedure and results we present in
Section~\ref{sec:spectral} are largely equivalent to those in
\citet{2014ApJ...797...92C}, who provide a much more detailed
description and in-depth analysis of the data.

We have also used a 2003 archival \xmm\ observation of \cen\ to
perform a deep search for coherent pulsations that would indicate a
relatively strong magnetic field (and suggest a magnetic
propeller). We describe this search in Section~\ref{sec:pulsations}.

\subsection{\hspace{-0.15cm}Spectral analysis}
\label{sec:spectral}
We analysed the \xmm\ European Photon Imaging Camera (EPIC;
\citealt{jansen2001,turner2001,strueder2001}) data of \cen\ (ObsID
0692790201) with the \textsc{sas} software, version 13.5.0, and the
latest available calibration files (in 2014 August). We reprocessed
the raw MOS and pn data with the \textsc{emproc} and \textsc{epproc}
tasks, in the latter case incorporating a correction for the effects
of X-ray loading on the pn data with \textsc{epxrlcorr}. We also
applied a correction for spatially dependent charge-transfer
inefficiency effects on the pn data using the \textsc{epspatialcti}
task. We searched for and excluded periods of flaring particle
background based on single-event light curves in the 10--15~keV band
for the MOS detectors and 10--12~keV band for the pn, using events
from the entire detector area in each case. To avoid the effects of
pile-up we extracted source spectra from annular regions. To decide on
appropriate excision radii we used the \textsc{epatplot} task. For
MOS1 and MOS2 we extracted source spectra from annuli with inner and
outer radii of 7.5 and 50 arcsec; background spectra were extracted
from source-free circular regions of radius 100 arcsec located on the
same chip as the source. For pn we extracted source counts from an
annulus between 10 and 40 arcsec; background counts were extracted
from a nearby chip, using a source-free 110 $\times$ 110 arcsec$^2$
rectangular region at a similar distance from the read-out node as the
source. We used standard event selection criteria. After creating
response files with \textsc{rmfgen} and \textsc{arfgen} we grouped the
spectra in the 0.3--10~keV range with the \textsc{specgroup} task,
imposing a minimum signal-to-noise ratio of 5, and limiting
oversampling of the energy resolutions of the detectors to a maximum
factor of 2.

We analysed the \nust\ \citep{harrison2013} data (ObsID 30001004002)
using the \nust\ Data Analysis Software (\textsc{nustardas}), version
1.4.1, in conjunction with the \nust\ calibration data base, version
20140715. We reprocessed the raw data using \textsc{nupipeline}, and
then extracted source spectra and created response files for both
(FPMA and FPMB) detectors with the \textsc{nuproducts} task. The
source spectra were extracted from circular regions of radius 75
arcsec centred on the source. Because of the significant variation of
the background across the two detectors, we modelled the background
contribution in the source extraction regions with the
\textsc{nuskybgd} tools package \citep{wik2014}. We grouped the
spectra in the 3--78~keV band, requiring a minimum signal-to-noise
ratio of 9 per group. This ensured that the energy resolution of the
detectors ($\sim$400 eV~below $\sim$50~keV) was nowhere oversampled by
more than a factor of 2.

We fitted the spectra from the five detectors (\xmm\ EPIC MOS1/MOS2/pn
in the 0.3--10~keV band and \nust\ FBMA/FPMB in the 3--78~keV band)
simultaneously with \textsc{xspec}, version 12.8.2. We initially
fitted the spectra with two additive components (one for the soft
thermal emission and another for the harder power-law component),
modified by absorption. In addition, we included in our model an
overall multiplicative factor -- fixed to 1 for the pn but free for
the other detectors -- to account for possible cross-calibration
shifts between the detectors. All other parameters were tied between
the five detectors. We used the \textsc{tbabs} absorption model with
\textsc{wilm} abundances \citep{wilms2000} and \textsc{vern}
cross-sections \citep{verner1996}. We modelled the soft component with
the \textsc{nsatmos} neutron star atmosphere model \citep{heinke2006},
fixing the neutron star mass at 1.4~$\textrm{M}_{{\sun}}$, the distance
at 1.2~kpc, and the fraction of the surface emitting at 1, but
allowing the effective surface temperature and the radius of the
neutron star to vary. Trying a fit with a simple power law for the
hard component [i.e. \textsc{const*tbabs*(nsatmos+powerlaw)}] we found
poor agreement with the data, with systematic residuals at the higher
energies indicating the presence of a break in the power law, as
reported by \citet{2014ApJ...797...92C}. We also found that replacing
the simple power law with a broken power law (\textsc{bknpower}), a
cut-off power law (\textsc{cutoffpl}), or a thermal bremsstrahlung
model (\textsc{bremss}) provided a significantly better fit for the
hard component, again in agreement with
\citet{2014ApJ...797...92C}. In Section~\ref{sec:RIAF} we describe how
we used these better-fitting two-component models as a basis for
placing limits on the possible presence of a second hard component in
the 0.3--78~keV spectrum.

\subsection{\hspace{-0.15cm}Search for pulsations from \cen}
\label{sec:pulsations}
If the magnetosphere is strong enough to truncate the disc and drive a
propeller or form a trapped disc, it is also strong enough to channel
the accretion flow on to the magnetic poles of the star, potentially
resulting in pulsed emission (as long as some material hits the
stellar surface; see Section \ref{sec:propeller}). If the emission
arises entirely from the surface of the star, any rotational asymmetry
in accretion on to the star should be detectable, since the accretion
flow itself is invisible. No pulsations have ever been detected in
\cen, but recent advances in pulse search techniques
\citep{2011PhRvD..84h3003M, 2012ApJ...744..105P} have made it possible
to search for high-frequency pulsations from very low luminosity
sources. In this spirit we have undertaken a systematic search for
pulsations in quiescent data from \cen\ using a semicoherent search
strategy \citep{2011PhRvD..84h3003M}.

\cen\ was observed in quiescence with \xmm\ on 2003 March 1 (MJD
52699) for a total on-source exposure time of ${\sim}80$~ks. The EPIC
pn detector operated in timing mode (with the thin filter), recording
data with a sampling time of $29.56$~$\umu$s. Since we searched
 spin periods down to the millisecond range, we selected only data coming
from the pn detector.

The data were processed using \textsc{sas}, version 13.0.0, with the
most up-to-date calibration files available when the reduction was
performed (2014 May). The photons were filtered by applying standard
screening criteria and by removing solar flares and telemetry
dropouts. After filtering, the total net exposure time was 68.5 ks.

To look for X-ray pulsations we adopted a semicoherent search
strategy \citep{2011PhRvD..84h3003M}. We define fully coherent
searches as those in which the phase of a matched filter tracks that
of the signal for the duration of the observation. For long
observations and large parameter spaces such searches rapidly become
computationally unfeasible. For finite computational power the most
sensitive searches are semicoherent \citep{2012PhRvD..85h4010P}. In a
semicoherent search the data are divided into short segments of
length $\Delta T$, and each segment is searched
\textit{coherently}. The different segments are then combined
incoherently. This is done in such a way as to ensure that signal
parameters are consistent between data products combined from
different segments, but does not impose signal phase consistency at
the segment boundaries. An extensive description of the technique can
be found in \citet{2011PhRvD..84h3003M} and in
\citet{2014arXiv1412.5938M} where the same scheme has been adopted to
search for pulsations in the LMXB Aql~X-1.

During the search we assumed a circular orbit and no prior knowledge
of the orbital phase, while we restricted the orbital period and
semimajor axis according to the values reported in
\citet{1989A&A...210..114C} (see also Table~\ref{tab:parspace}). To
avoid the influence of possible unreported systematics we extended the
range of values used for the orbital period. While the search is
sensitive to signals with time-varying amplitude (on time-scales
$>$$\Delta T)$, for the purposes of our reported sensitivity we assume a
constant fractional amplitude of the pulsations for the entire
duration of the observation.

\begin{table}
\begin{center}
\label{tab:parspace}
\begin{tabular}{@{}cccc@{}}
\multicolumn{4}{@{}p{4.5cm}@{}}{\textbf{Table 1.} Parameter space boundaries for the \cen\ search.}\\[5mm]
\hline
Parameter & Units & Min & Max \\ 
\hline
$\nu$ & Hz & 50 & 1500\\
$a_1$ & lt s & 0.04 & 1.9 \\
$P_{\rm b}$ & s & 54\,000 & 54\,720 \\
$T_{\mathrm{asc}}$ & s & \multicolumn{2}{c}{Full orbit}\\
\hline\\[-4.3mm]
\multicolumn{4}{@{}p{4.5cm}@{}}{{\it Note.} $\nu$ is the spin frequency; $a_1$ is the
projected semimajor axis of the neutron star orbit; $P_{\rm b}$ is the
orbital period; and $T_{\mathrm{asc}}$ is the time of passage through
the ascending node (corresponding to the point at $0\degr$ true
longitude).}\\
\end{tabular}
\end{center}
\end{table}

The data were divided into $443$ segments of $\Delta T=128$~s that
were each searched coherently.  A cubic grid of templates using a 10
per cent worst-case mismatch was then placed on the two-dimensional
parameter space\footnote{Within each segment the signal was
  approximated by its second-order Taylor expansion in phase. The
  choice of maximum expansion coefficient is dictated by the choice of
  fixed segment length (see \citealt{2011PhRvD..84h3003M}).} of the
spin and spin derivative. To combine the data products from each
segment an additional bank of random templates was then used to cover
the space defined by the Cartesian product of the physical parameter
ranges given in Table~\ref{tab:parspace}. This bank was constructed so
as to provide a 90 per cent coverage at 10 per cent mismatch, and
consequently contains $3.42\times 10^{10}$ trials.

Our expected theoretical sensitivity is 6.8 per cent rms, adopting a
multitrial false-alarm rate of 1 per cent (conservatively assuming
that our trials are independent) and a 10 per cent false-dismissal
probability. We detected no significant pulsations with a
fractional amplitude upper limit of 6.4 per cent.

\section{\hspace{-0.1cm}\an{The origin of the quiescent Cen~X-4 spectrum}}
\label{sec:phys_proc}
What is the origin of the power-law component in the quiescent 
X-ray spectrum of \cen? While the soft component is well described by
quasi-thermal emission from the entire neutron star surface
\citep{1999ApJ...514..945R,2010ApJ...720.1325C}, the origin of the
power law is unclear. Its luminosity clearly comes from accretion
energy, but the dominant radiation mechanism, geometry, and location
of the emission are uncertain. As we show below, the low-energy
spectral cut-off and hard spectral index clearly suggest
bremsstrahlung emission (as previously concluded by
\citealt{2014ApJ...797...92C}), while the near-balance between the
thermal and power-law components strongly suggests the power law is
emitted from the boundary layer of the star, rather than in the
accretion flow.

\subsection{\hspace{-0.15cm}Emission from the accretion flow}
\label{sec:chakra}

\subsubsection{\hspace{-0.15cm}Constraining radiation from a radiatively inefficient flow}
\cite{2014ApJ...797...92C} interpret the power-law X-ray spectral
component of \cen\ as an ADIOS flow -- a radiatively inefficient
flow model proposed by \cite{1999MNRAS.303L...1B}. In this model, the
inner $10^3$--$10^5 R_{\rm S}$ of the accretion flow no longer forms a
thin disc, and instead gravitational potential energy released by
infall is primarily advected outwards, launching a strong wind from
the outer part of the accretion flow. As a result the net accretion
rate in this region becomes a strong function of radius:
\beq
\dot{M}(r) = \dot{M}(r_{\rm in})\left(\frac{r}{r_{\rm in}}\right)^p,
\eeq
where $ 0 \leq p < 1$ is a scaling parameter for the local accretion
rate as a function of radius: $p = 0$ gives the standard solution
(with no outflow), while $p < 1$ is imposed so that the amount of
gravitational energy released increases with decreasing radius. The
accretion rate through the flow can thus vary by several orders of
magnitude between the inner and outer edges of the ADIOS flow, $r_{\rm in}$ and
$r_{\rm out}$.

Fitting the hard X-ray spectrum with a thermal bremsstrahlung model,
\cite{2014ApJ...797...92C} find a gas temperature of $18$~keV. In
their interpretation of the spectrum, they assume the entire accretion
flow has this temperature and use the observed luminosity to infer a
density profile for the ADIOS flow. They assume that the total
luminosity from the source corresponds to the net accretion rate on to
the star at $r_{\rm in} \sim R_{\rm *}$ (see equation \ref{eq:edot})
so that $\dot{M} \simeq 4\times10^{-6}$~$\dot{M}_{\rm Edd}$. They adopt a
thick toroidal geometry for the flow with a radial velocity $v_{\rm r}
= \mu \sqrt{GM_*/r}$ (where $\mu \simeq 0.1$), meaning
that the infall velocity is a considerable fraction of the Keplerian
velocity. The hardness of the spectral index and low energy of the
spectral cut-off exclude synchrotron emission and inverse Compton
scattering as plausible emission mechanisms, and
\cite{2014ApJ...797...92C} conclude that the hard X-rays are most likely
produced by bremsstrahlung emission. They estimate $p >$~0.8--1 for the
accretion rate as a function of radius.

Overall, this interpretation presents several difficulties in
explaining the behaviour of \cen. First, it essentially assumes that
the flow is radiatively efficient: the amount of luminosity in the
flow is roughly equal to the net accretion rate at $r_{\rm in}$. The
radiative efficiency of an ADIOS flow is not defined a priori,
and is so high in the present case because there is so much hot gas in
the outer parts of the flow. With a temperature of 18~keV, the gas
around $r_{\rm out}$ has a sound speed, $c_{\rm s}$, equal to or
considerably greater than the Keplerian velocity of the flow, $v_{\rm
  K}$ (and considerably larger than the virial velocity, about $0.4
v_{\rm K}$; \citealt{1995ApJ...452..710N}). In contrast, the
\cite{1999MNRAS.303L...1B} model predicts a sound speed of at most
$c_{\rm s} \sim 0.6$--$0.8\,v_{\rm K}$. This mismatch is only
exacerbated by the observation that the spectral index decreases at
lower luminosities \citep{ 2010ApJ...720.1325C, 2014ApJ...797...92C},
which \cite{2014ApJ...797...92C} interpret as evidence of an increased
gas temperature with decreasing $\dot{M}$.

As mentioned previously, if the power-law component is emitted by the
accretion flow, the approximately equal contribution of the thermal
and power-law components of the spectrum is not naturally explainable,
nor are the short time-scales on which they covary. As presented in
\cite{1999MNRAS.303L...1B}, and recently revisited by
\cite{2012MNRAS.420.2912B}, the ADIOS flow can be extremely
radiatively inefficient (i.e. adiabatic): the amount of radiation from
the flow can be arbitrarily small, since most of the accretion energy
is used to launch an outflow. The observed balance between the thermal
and power-law components would therefore be coincidental. The large
geometric separation between the thermal component (at $R_*$) and
power-law component (at $\sim$$10^4\,R_*$) is also difficult to reconcile
with the rapid observed variability. At $10^4\,R_*$ the characteristic
accretion time, $t_{\rm visc} \sim r/v_{\rm r} = \mu^{-1}
\sqrt{r^3/GM_*}$, is about 770 s, or nearly an order of magnitude
larger than the shortest observed time-scale for covariation of the
two spectral components \citep{2004ApJ...601..474C,
  2010ApJ...720.1325C, 2014ApJ...797...92C}.

The considerable uncertainty in the accretion rate at $r_{\rm in}$ can
alleviate the energetic difficulties of the proposed ADIOS solution.
\cen\ could be in a strong propeller state, so that most of the
accretion flow is expelled by the magnetic field when it reaches \rin\
(although as we argue in Section \ref{sec:magnetosphere} the lack of
detected pulsations makes this seem unlikely). The strength of the
thermal component is set by residual gas accreting on to the star,
which could then be much lower than the accretion rate in the inner part of
the accretion flow. If the accretion rate in the flow is considerably
higher and most of the gas is expelled in a propeller outflow, the gas
density and temperature will be higher, so that a weaker outflow and
smaller ADIOS region will produce the observed emission. However, if
there is a strong propeller the covariation of the thermal and
power-law components and their near balance is hard to understand,
since now a strong propeller will further break the connection between
accretion flow and the surface of the star. Although the ADIOS model
(and other radiatively inefficient accretion flows) can qualitatively
explain the observed power-law emission from \cen, we conclude that
the model is implausible in detail.

\subsubsection{\hspace{-0.15cm}Emission from a radiatively efficient flow?}
\label{sec:inner}
Simple physical arguments can also be used to show that the emission
cannot be produced by a radiatively efficient flow in the inner part
of the flow. A geometrically thin, optically thin hot accretion flow
is unstable to expansion since it cannot cool radiatively as fast as
it is heated by turbulence \citep{2002apa..book.....F}. This expansion
implies that at the low observed accretion rates the gas density would
not be high enough to produce the observed spectrum from
bremsstrahlung (which would require a very small disc scale height). In
black holes at somewhat higher accretion rates, it has been suggested
that the hard X-ray emission comes from an optically thin corona
overlying a cold, optically thick disc, although in this case the
underlying disc is heated and should produce comparable luminosity to
the corona \citep{1991ApJ...380L..51H} and a strong Fe K emission line
(e.g. \citealt{1993MNRAS.261...74R}), neither of which are
observed. Since roughly 50 per cent of the emission is coming from the
surface, the energy balance between the hard and soft component is not
accounted for: the underlying disc should be as luminous as the
corona, and the blackbody component (made up of a disc and the
surface) should clearly dominate over the power-law one.

A simple energy balance calculation shows that the emission cannot
come from inverse Compton scattering of a large optical depth corona
above a thin disc. \cite{2014ApJ...797...92C} fit the power-law
component with an inverse Compton spectrum from a disc, and find $\tau
\sim 4$ and $kT \sim 6$~keV. The scale height of such a corona is 
\beq
H = \left(\frac{kTr}{m_{\rm p} GM_*}\right)^{0.5} r\sim 0.01r, 
\eeq
which implies a scale height of about $10^4$ cm in the inner parts of
the flow.

For an electron-scattering optical depth of $\tau = \sigma_{\rm T}
n_{\rm e}H\sim 4$ (assuming ionized hydrogen gas), the surface density of the
corona will be
\beq
H n_{\rm e} \simeq 6 \times 10^{24} {\rm~cm^{-2}}.
\eeq
This implies a number density of $n_{\rm e} \sim 6\times 10^{20} {\rm
 ~cm^{-3}}$ in the inner accretion flow, which is too high for the
heating rate (from turbulent viscosity) to balance radiative losses
from bremsstrahlung at the observed gas temperature, as a simple
calculation shows. Assuming $\dot{M} = 8.1\times10^{12} {~\rm
  g~s^{-1}}$ (from the observed luminosity), the local heating rate in
the inner disc will be $\sim$$3GM_*\dot{M}/(8\upi R^3_*) \sim
9.6\times10^{19} {~\rm erg~s^{-1}~cm^{-2}}$
\citep{2002apa..book.....F}. The frequency-integrated bremsstrahlung
emission function (erg s$^{-1}$ cm$^{-3}$) for ionized gas of density
$n_{\rm e}$ and temperature $T_{\rm e}$ is (assuming ionized hydrogen and a Gaunt
factor of 1.2; \citealt{1986rpa..book.....R})
\beq
\label{eq:brems}
U_{\rm brems} = 1.7\times10^{-27} n_{\rm e}^2T_{\rm e}^{1/2}.  \eeq
The emission rate per unit area of the corona (integrated over the
scale height of 10$^4$~cm) will thus be $\sim$$3\times 10^{23} {\rm
  ~erg~s^{-1}~cm^{-2}}$ which is much larger than the energy released
in turbulent processes.

There is thus no obvious way for an accretion flow (radiatively
efficient or not) at such a low accretion rate to produce the
(apparent) bremsstrahlung emission unless the accretion rate at the
inner edge is considerably larger than implied by the observed
emission. This requires a strong propeller, which then does not
explain the apparent balance between the thermal and power-law
components, nor the observed covariation of the two components on
short time-scales.

\subsection{\hspace{-0.15cm}Emission from a boundary layer}
\label{sec:boundary}
Given that more than half the emission in \cen\ comes from a quasi-blackbody
component mostly generated by accretion, and that the two spectral
components are roughly equal and covary, the power-law component
could also plausibly originate close to the neutron star
surface.\footnote{In the refereed version of their work,
  \cite{2014ApJ...797...92C} have independently suggested the same
  possibility, and reach similar conclusions to ours as to the nature
  of the boundary layer.} As discussed in Section \ref{sec:models}, a
large fraction (at least 1/4--1/2) of the total accretion energy is
released in the boundary layer of the star, in the final impact of gas
on to the stellar surface. Although it is not clear how the gas's
kinetic energy is converted to radiation in the boundary layer, it is
reasonable to assume that not all the radiation will come from an
optically thick layer and there may be a strong power-law
contribution. 

The most plausible radiation mechanism to explain the power-law
spectrum is optically thin thermal bremsstrahlung. The joint \xmm\ and
\nust\ fit of \cite{2014ApJ...797...92C} suggests the bremsstrahlung
component has a luminosity of $L_{\rm X} \simeq
7.2\times10^{32}$~erg~s$^{-1}$ (0.3--78 keV, assuming a distance of 1.2~kpc) and temperature of $kT_{\rm e}\simeq 18$~keV for the bremsstrahlung
component. The total inferred luminosity from 0.3 to 78 keV is $1.5\times
10^{33}$~erg~s$^{-1}$.  Inverse Compton scattering from a surface
corona is excluded, since if the optical depth were high enough to
produce the emission it would completely obscure the (thermal) surface
emission from the star. Synchrotron emission may also be important if
the boundary layer has magnetic turbulence with $\beta = P_{\rm
  gas}/P_{\rm B} \sim 1$, but again the low cut-off energy is
inconsistent with emission arising from power-law electrons
accelerated by shocks (which in general has a much higher cut-off
energy). A model which produces synchrotron emission from the
interaction of a pulsar radio jet with the outer accretion disc
(suggested by \citealt{2004ApJ...601..474C}) is carefully considered
and disfavoured in \cite{2014ApJ...797...92C}.

If the boundary layer covers the entire surface of the star with a
scale height $b$ and emits bremsstrahlung radiation, the amount of
bremsstrahlung radiation \bhl and its temperature (inferred from the strength
and cut-off energy of the X-ray spectral power-law component) can be
used to relate the mean \ehl density and
scale height of the boundary layer:
\begin{align}
  L_{\rm brems} \mspace{1.5mu} &=  \mspace{1.5mu} U_{\rm brems} 4\upi R_*^2 b\nonumber\\
   &= \mspace{1.5mu} 3.1\times10^{-10}~n_{\rm e}^2 b \left(\frac{kT_{\rm e}}{18\,\textrm{keV}}\right)^{1/2}\left(\frac{R_*}{10^6\,\textrm{cm}}\right)^2.
\end{align}
Thus,
\begin{align}
\label{eq:n2b}
n_{\rm e}^2b \mspace{1.5mu} &= \mspace{1.5mu} 2.3\times10^{42}\left(\frac{R_*}{10^6\,\textrm{cm}}\right)^{-2}\nonumber\\
 & \mspace{20mu} \times\left(\frac{L_{\rm brems}}{7.2\times10^{32}\,\textrm{erg\,s}^{-1}}\right)
\left(\frac{kT_{\rm e}}{18\,\textrm{keV}}\right)^{-1/2} {\rm~cm^{-5}}.
\end{align}

We can set an upper limit on the density of the boundary layer by
considering the accretion rate required for the gravitational
potential energy released to match the observed luminosity. Assuming
the entire luminosity ($L_{\rm tot}$) comes from the boundary layer,
the accretion rate on to the star will be 
\beq
\label{eq:mdotL}
\dot{M} = \frac{L_{\rm
    tot}}{v^2_{\rm K,*}\mathcal{F}_{\rm bl}}, 
\eeq
 where $v_{\rm K,*}
\equiv (GM_*R_*^{-1})^{1/2}$ is the Keplerian velocity at $R_*$ and
\fbl\ is the fraction of gravitational potential energy released in
the boundary layer. 

The accretion rate on to the star will be
\beq
\label{eq:mdot}
\dot{M} = 4\upi R_*^2 m_{\rm p} n_{\rm e} v_{\rm R,*},
\eeq
where $v_{\rm R,*}$ is the radial infall velocity in the boundary
layer. Combining equations (\ref{eq:mdot}) and (\ref{eq:mdotL}) gives the
characteristic number density in the boundary layer:
\begin{align}
\label{eq:ne}
n_{\rm e} \mspace{1.5mu} &= \mspace{1.5mu} 7.7\times10^{23} \left(\frac{L_{\rm
      tot}}{1.5\times10^{33}\,\textrm{erg\,s}^{-1}}\right) \left(\frac{\mathcal{F}_{\rm bl}}{0.5}\right)^{-1}\nonumber\\
 & \mspace{20mu} \times\left(\frac{M_*}{1.4\,\textrm{M}_{\sun}}\right)^{-1}\left(\frac{R_*}{10^6\,\textrm{cm}}\right)^{-1} v_{\rm R,*}^{-1}{\rm~cm^{-3}}.
\end{align}
 The boundary layer (assuming there is no magnetosphere) is expected to
be strongly turbulent, since infalling gas must transition from
roughly Keplerian orbits to corotation with the
star. \cite{1988AdSpR...8..135S} derived some simple physical
properties of the boundary layer, assuming that a turbulent flow
develops. The net infall velocity of the gas can be expressed as  
\beq
v_{\rm R,*} \simeq v_{\rm K,*} \frac{\mathcal{M}^2 (3 + \Omega_*/\Omega_{\rm K,*})}{2C}\frac{b}{R_*}, 
\eeq
 where
$\mathcal{M}$ is the Mach number of the flow, $\Omega_*/\Omega_{\rm
  K,*}$ is the ratio of the stellar spin frequency to the Keplerian
frequency at the surface of the star, and $C$ is a numerical factor of
order unity (representing uncertainty in the nature of the
turbulence). If we assume strong, nearly supersonic turbulence with
$\mathcal{M} \sim 1$, the maximum infall velocity  will be
\beq 
v_{\rm R,*} \sim \frac{3}{2}\frac{b}{R_*}v_{\rm K, *}.  
\eeq 
This sets an upper limit on $n_{\rm e}b$,
\beq
n_{\rm e}b > 3.8\times10^{19}\left(\frac{R_*}{10^{6}\,\textrm{cm}}\right)^{1/2}.
\eeq
From equation (\ref{eq:n2b}) we thus get an upper limit on the number
density of the boundary layer:
\beq
n_{\rm e} < 6\times10^{23}{\rm ~cm}^{-3}.
\eeq
We can also infer a hydrostatic scale height for the corona. For a
coronal temperature of 18~keV (corresponding to a sound speed $c_{\rm
  s} = \sqrt{kT/m_{\rm p}} \simeq 1.3\times10^8 {\rm~cm~s^{-1}}$) the
scale height will be:
\beq 
b \sim R_* \left(\frac{c_{\rm s}}{v_{\rm
      K,*}}\right)^2 \sim 10^2 { \rm~cm}, 
\eeq which is consistent
with the limits we have derived and implies a number density of
$n_{\rm e} \sim 6\times10^{19} {\rm~cm^{-3}}$ (note that this implies
a Compton optical depth $\tau \sim \sigma_{\rm T}n_{\rm e}b <
10^{-2}$, so that electron scattering will essentially be irrelevant).

This solution assumes that the boundary layer is a single-temperature
gas. This will be true as long as the density is high enough for the
ions and electrons to maintain collisional equilibrium faster than the
ions can heat (via turbulence) or the electrons can cool by radiating.
As a final check on our solution above, we can compare the predicted
turbulent heating time-scale (the infall time), the cooling time-scale
(bremsstrahlung emission from electrons), and the time-scale for the
electrons and ions to reach collisional equilibrium.

The characteristic heating time for the gas is approximately the
infall time:
\beq
t_{\rm heat} \simeq \frac{b}{v_{\rm R,*}}  \simeq \Omega_{\rm K,*}^{-1} =
7.3\times10^{-5} {\rm~s},
\eeq
while the time-scale for the gas to cool via bremsstrahlung emission
will be (e.g. \citealt{1987A&A...185..155K})
\begin{align}
t_{\rm brems} \mspace{1.5mu} &\simeq \mspace{1.5mu} 2\times10^{11}n_{\rm e}^{-1}T_{\rm e}^{1/2}\\
 \mspace{1.5mu} &= \mspace{1.5mu} 5.7\times10^{-5}\left(\frac{n_{\rm e}}{6\times10^{19}\,\textrm{cm}^{-3}}\right)^{-1}\left(\frac{kT_{\rm e}}{18\,\textrm{keV}}\right)^{1/2} {\rm~s}.\nonumber
\end{align}
The time-scale for electrons and ions to reach equilibrium will be
\citep{1969pldy.book.....B}
\beq
  t_{\rm e-i} = 27.5 T_{\rm e}^{3/2} n_{\rm e}^{-1}\left(1 +
    \frac{1}{\xi}\right),
\eeq 
where $\xi \equiv (T_{\rm i} - T_{\rm e})/T_{\rm e}$ is the
fractional difference between the ion and electron temperatures. For 
$n_{\rm e} \sim 6\times10^{19} { \rm
~cm}^{-3}$, and the inferred gas temperature, $18$~keV,
$t_{\rm e-i} \sim t_{\rm heat}$ even for a significantly two-temperature plasma:
\begin{align}
t_{\rm e-i} \mspace{1.5mu} &<  \mspace{1.5mu} 1.4\times10^{-6}\left(1 + \frac{1}{\xi}\right) \left(\frac{n_{\rm e}}{6\times10^{19}\,\textrm{cm}^{-3}}\right)^{-1}\nonumber \\
& \mspace{20mu} \times \left(\frac{kT_{\rm e}}{18\,\textrm{keV}}\right)^{3/2}{\rm~s}.
\end{align}
Furthermore, for a small boundary layer scale height,
the radiative cooling time is significantly longer than the cooling
time from expansion: 
\begin{align}
  t_{\rm exp} \mspace{1.5mu} &\sim \mspace{1.5mu} \frac{b}{c_{\rm s}}\nonumber\\
   \mspace{1.5mu} &\simeq \mspace{1.5mu} 8\times10^{-7} \left(\frac{b}{10^2\,\textrm{cm}}\right)\left(\frac{kT_{\rm e}}{18\,\textrm{keV}}\right)^{-1/2}{\rm~s}.
\end{align}

\bhl The coronal layer may therefore expand into a two-temperature plasma
with a significantly larger scale height than $b\sim10^{2}$ cm, but a
more detailed energy balance is needed to explore the coronal
structure in more depth. Regardless of this uncertainty,\ehl~the basic
model for bremsstrahlung from a surface corona fits the observed
spectrum for a reasonable range of scale heights and densities and
satisfies basic energy balance requirements. It also offers a
reasonable explanation for the energy balance between the
bremsstrahlung and blackbody components.


\subsection{\hspace{-0.15cm}How radiatively inefficient is the accretion flow?}
\label{sec:RIAF}
The quiescent X-ray spectrum of \cen\ is most likely produced by the
final fall of accreting matter on to the surface of the neutron
star. The accretion flow itself (which should radiate about half
the total liberated gravitational energy if the flow is efficiently
radiating; Section \ref{sec:models}) is not detected, and we can
constrain its intrinsic radiative efficiency from setting upper limits on the
presence of a second hard X-ray component (with luminosity $L_{\rm
  flow}$) in the observed X-ray spectrum. From the upper limit on
$L_{\rm flow}$ we can use the limits on the energy available to be
radiated from each component obtained in Section~\ref{sec:models} to set
upper limits on the intrinsic radiative efficiency of the accretion
flow, assuming it radiates primarily in X-rays.

To estimate a rough upper limit on the possible flux from the
accretion flow \bhl we added a third emission component to the three better-fitting two-component models \ehl\ discussed in
Section~\ref{sec:spectral} (i.e. the ones where the hard component was
modelled by a broken power law, a cut-off power law, or a thermal
bremsstrahlung model). \bhl We initially used a simple (unbroken) power
law (\textsc{pegpwrlw}) for this extra component\ehl, since this model
has only two parameters (a slope and normalization), and since
power-law emission is typically seen in the accretion flow of both
black hole and neutron star binaries at low luminosities. We defined
the normalization of this additional power-law component to be its
unabsorbed flux in the 0.3--78~keV band (i.e. the combined \xmm\ and
\nust\ range). For a given fixed photon index we then derived a
95 per cent confidence upper limit on this power-law flux (using
$\Delta\chi^2=2.706$) and converted this upper flux limit to an upper
limit on the fractional contribution of the power-law component to the
total 0.3--78~keV unabsorbed flux. We explored the range 1.0--2.5 for
the photon index of the additional power law. We carried out this
procedure for each of the three original hard-component models
(\textsc{bknpower}, \textsc{cutoffpl}, and \textsc{bremss}). In the
specific cases of the broken power law and cut-off power law we
constrained the values of the photon indices of those components to be
$\geq$1.0 to prevent the indices from assuming unphysically low values
during the upper limit calculation.

The procedure described above yielded 95 per cent upper limits on the
fractional contribution of the extra power-law component to the total
unabsorbed 0.3--78~keV flux in the ranges $\sim$0.11--0.34,
0.06--0.33, and 0.06--0.26, for the models with a broken power law, a
cut-off power law, and thermal bremsstrahlung, respectively, when
varying the photon index of the additional power law between
1.0 and 2.5. In all three cases the highest upper limits were obtained when
the photon index of the extra power law was in the range
$\sim$1.5--1.7. We also repeated our analysis with the neutron star
mass fixed at 1.9 $\textrm{M}_{{\sun}}$ (the value used by
\citealt{2014ApJ...797...92C}) instead of 1.4 $\textrm{M}_{{\sun}}$,
but found that this change had a negligible effect on our results.

\bhl The exact shape of a possible extra spectral component arising
from the accretion flow itself is uncertain; e.g. it is possible that
rather than being a simple power law throughout the 0.3--78 keV range
there could be a low- and/or high-energy roll-over present that
affects the emission in the observed bandpass. Therefore, to make the
analysis more robust, we repeated our analysis procedure with the
simple power law replaced by the \textsc{nthcomp} Comptonization model
\citep{zdziarski1996,zycki1999}. We explored a range of values for
three of the model parameters. For the seed photon temperature
(governing the low-energy roll-over) we explored the range 0.2--1.0
keV; for the electron temperature (governing the high-energy
roll-over) we considered values of 5--300 keV; and for the asymptotic
power-law photon index (set by the optical depth of the Comptonizing
plasma as well as the electron temperature) we considered the range
1.1--2.5. The seed photon distribution was assumed to have a blackbody
shape. Unsurprisingly, it was possible to find combinations of
parameter values that allowed a somewhat larger contribution from this
component than was the case for a simple power law; on the fractional
contribution of the \textsc{nthcomp} component to the total unabsorbed
0.3--78 keV flux we find 95 per cent upper limits that range from
$\sim$0.03 to as much as $\sim$0.46.

For the largest fractional contributions, however, the additional
\textsc{nthcomp} component essentially replaces the original hard
component with one that is physically implausible. The
\textsc{nthcomp} component in these cases has a seed photon
temperature of $\sim$0.7--1 keV and a typical photon index of
$\sim$1.9--2.0. The fractional contribution of the component is
virtually independent of the upscattering electron temperature (which
determines the hard X-ray shape of the spectrum), indicating that the
contribution to the soft X-rays is what chiefly limits the weight of
the additional component. In this fit, the X-ray spectrum up to the
high-energy roll-over is essentially fitted by the peak of the
blackbody seed photon distribution with some upscattered photons,
while the observed break is reproduced by the steeper power-law tail
of the Comptonized photon distribution. There is no independent
indication or theoretical expectation for such a hard X-ray seed
photon population, so the addition of a more complex model mainly
serves to demonstrate that the detailed shape of an additional
component in the hard X-rays does not significantly influence how
strong the contribution of such a component could be. We thus use the
limits given by the addition of the simpler model (a power law) to set
limits on the intrinsic radiative efficiency of the accretion
flow. \ehl

We conclude that the luminosity of the accretion flow itself, $L_{\rm
  flow}$, can make up no more than $\sim$6--34 per cent of the total X-ray
luminosity, and an even smaller fraction of the total luminosity,
given the significant observed UV flux. 
Since the UV contribution and origin are
still unclear, however, we do not include it in our estimate of
$L_{\rm flow}/L_{\rm tot}$.

The luminosity of a given spectral component will be \beq L =
\dot{E}_{\rm pot}\mathcal{F}\varepsilon, \eeq where $\varepsilon$ is
the intrinsic radiative efficiency of the component (i.e. the fraction
of the gravitational potential energy available to it that the
component radiates away). The system's total luminosity will then be:
\beq L_{\rm tot} = L_{\rm flow} + L_{\rm bl} = \dot{E}_{\rm
  pot}(\mathcal{F}_{\rm flow}\varepsilon_{\rm flow} + \mathcal{F}_{\rm
  bl}\varepsilon_{\rm bl}).  \eeq By measuring $L_{\rm tot}$, assuming
$\varepsilon_{\rm bl} = 1$ (the boundary layer radiates energy
efficiently), and using our upper limits on $L_{\rm flow}$ (see above)
we can estimate $\varepsilon_{\rm flow}$, provided we can constrain
\fbl\ and \fflow. As described in Section \ref{sec:models}, this will
depend on the neutron star's spin, size, and magnetic field, as well
as the width of the boundary layer. Since the gravitational potential
energy liberated by the different components can be unequal,
i.e. the intrinsic efficiency can still be high if there is very
little energy available to be radiated in the flow compared with the
disc (as in e.g. a trapped disc scenario). The intrinsic radiative
efficiency of the accretion flow is then \beq \varepsilon_{\rm flow}
= \frac{\mathcal{F}_{\rm bl}}{\mathcal{F}_{\rm flow}} \frac{L_{\rm
    flow}}{L_{\rm bl}}=\frac{\mathcal{F}_{\rm bl}}{\mathcal{F}_{\rm
    flow}} \left(\frac{L_{\rm flow}/L_{\rm tot}}{1 - L_{\rm
      flow}/L_{\rm tot}}\right). \eeq For our observed range of upper
limits on $L_{\rm flow}$, this gives $\varepsilon_{\rm flow} <
0.06$--0.5 if \fbl~=~\fflow. If none of the factors below plays a
significant role, this demonstrates that the flow is intrinsically
radiatively inefficient. As elsewhere in the paper, we assume that the
boundary layer is radially narrow ($b \ll R_*$), based on the
reasoning presented at the beginning of Section \ref{sec:models}.

The size, spin, and magnetic field of \cen\ are all unknown. However,
the absence of pulsations and presence of Type I X-ray bursts suggest
a relatively weak magnetic field (we adopt $10^{8} {\rm~G}$ as a
fiducial magnetic field for the calculations below), while the large
observed outbursts (and hence high mass transfer rates) suggest a
large amount of spin-up and a fast spin period, so we adopt 2 ms as a
fiducial spin period.

From Section \ref{sec:models}, very small (soft equation of state)
slowly rotating neutron stars (over a range of mass) can have
$\max(\mathcal{F}_{\rm bl}/\mathcal{F}_{\rm flow}) \sim 1.8$--2, which
corresponds to a radiative efficiency $\varepsilon_{\rm flow}~<~ 0.1$--$1$
(considering the range of upper limits on $L_{\rm
  flow}$). 

Alternatively, if the star is rapidly rotating without a large
magnetosphere (the most likely scenario, given the lack of any
indication of a magnetic field) the boundary layer will be dimmer than
the accretion flow, and $\mathcal{F}_{\rm bl}/\mathcal{F}_{\rm flow}
\sim 0.4$--0.6, which will correspond to a smaller intrinsic
efficiency for the flow: $\varepsilon_{\rm flow} < 0.03$--0.3.

If a magnetic field does regulate the flow near the star the two
scenarios presented in Section \ref{sec:magnetosphere} give
significantly different predictions for $\varepsilon_{\rm
  flow}$. Using the propeller and trapped disc models presented in
Fig.~\ref{fig:fbl} we can estimate $\dot{M}$, \fbl, and
\fflow\ that correspond to the observed luminosity,
$ L_{\rm X} \simeq 9 \times 10^{-6} L_{\rm Edd}$.

If a strong propeller is operating, the observed boundary layer
luminosity corresponds to an accretion rate of $\dot{M} =
1.4\times10^{-4} \dot{M}_{\rm Edd}$, with $\mathcal{F}_{\rm bl} =
0.04$ (using the fit to the MHD simulations of
\citealt{2006ApJ...646..304U}). The corresponding energy release in
the accretion flow is $\mathcal{F}_{\rm flow} = 0.09$. This
corresponds to a very small intrinsic efficiency for the flow,
$\varepsilon_{\rm flow} < 0.03$--0.2.

Alternatively, if a trapped disc forms, the flow is truncated close to
the magnetospheric radius (around 30 km) and is thus much less
luminous than the boundary layer: $\mathcal{F}_{\rm flow} \simeq 0.17 \mathcal{F}_{\rm bl}$. In
this case the intrinsic radiative efficiency of the accretion flow is
difficult to constrain, and we find $\varepsilon_{\rm flow} \sim 0.4$--1 as an upper
limit. Note that the `trapped disc' presented here is an extreme
limiting case: in reality there may be some outflow (even substantial)
where the flow hits the magnetic field. As long as there is
considerable angular momentum from the star transported through the
inner radius of the flow, the disc will stay trapped and $r_{\rm m}$
will remain close to $r_{\rm c}$. Any outflow of matter in the inner
parts of the flow will decrease the radiative efficiency of the
boundary layer, and thus require a smaller value for $\varepsilon_{\rm
  flow}$.

For most reasonable values of the neutron star spin, size, and
magnetic field, the non-detection of an accretion flow component in
\cen\ thus demonstrates that the flow is intrinsically radiatively
inefficient. \cen\ shows no indication of a significant magnetic field
(from our pulsation search), so we consider the most likely scenario
(based on the known history and properties of the star) to be a
rapidly spinning weakly magnetized star. In this case the intrinsic
efficiency is $\lesssim$$30$ per cent. In the least constraining case (a
trapped disc with no magnetically driven outflow) the limits cannot
conclusively show a radiatively inefficient flow, but again this case
may represent an extreme solution.

The hard X-ray spectrum with a cut-off at $\sim$10~keV, the balance
between the quasi-blackbody and power-law components and the
covariation of the two all strongly suggest that the power-law
emission originates close to the star, most likely via accretion on to
the surface of the star. By extension, the power-law emission seen in
other quiescent neutron star binaries likely has the same origin, and
their spectra could be interpreted in the same way. Moreover, an
observation during the decay phase of an outburst could allow both the
emission from the boundary layer and the accretion flow to be detected
simultaneously as power-law components with two different cut-off
energies. The hard X-ray sensitivity of \nust\ might be able to
detect the cut-off of one component directly, which would allow a
direct measurement of the strength of each component, and thus offer
better constraints on RIAF models. Repeating the analysis above on a
neutron star with a better constrained spin period and magnetic field
would allow better limits on \fbl\ and \fflow, although as our
analysis above demonstrates, the main uncertainty in \fbl\ and \fflow\
is for the magnetic case where $r_{\rm m} > r_{\rm c}$.

\section{\hspace{-0.1cm}\an{Discussion}}
\label{sec:discussion}
In this paper we have argued that the quiescent emission from the
neutron star \cen\ originates predominantly from the impact of the
accretion flow on to the surface of the star (most likely without
strong modulation by the magnetic field), without an equivalent contribution from
the accretion flow. This conclusion has consequences for both the
properties of the accretion flow (beyond its intrinsic radiative
inefficiency) and the neutron star itself, and suggests future
observations that can better constrain the properties of both.

Unless there is a very strong magnetic propeller (disfavoured by the
lack of pulsations), the low luminosity of \cen\ disfavours a strongly
advecting accretion flow (ADAF), since in an ADAF the accretion rate
is moderately large ($\sim$$ 10^{-3}$--$10^{-2}$$\dot{M}_{\rm Edd}$)
and the majority of the accretion energy is carried on to the central
object. This conclusion is the same as was reached by
\cite{2013MNRAS.436.2465B} and opposite to that of
\cite{2001ApJ...557..304M}, who interpreted the low surface emission
from \cen\ as evidence of a strong magnetic propeller. As we have
demonstrated in this paper, \cen\ shows no evidence for magnetic
activity: our pulse search did not find pulsations (with an upper
limit of 6.4 per cent pulsed fraction), which are likely to be present
if a strong magnetic field is channelling the flow (via episodic
accretion, see e.g. Section \ref{sec:propeller}). A more plausible
scenario is that the majority of the gas in the accretion flow does
not reach the star, as is proposed by various RIAF models
(e.g. \citealt{1999MNRAS.303L...1B,2000ApJ...539..809Q,2003PASJ...55L..69N}).

Various RIAF models generally require substantially more gas in the
accretion flow than is accreted on to the star, and this gas could
obscure the stellar surface if the density is high enough and the
orbital inclination angle is high (i.e. the accretion flow is observed
\bhl close to edge on as in dipping neutron star binary sources; see
e.g. \citealt{2012ApJ...760L..30H} for a list of such
systems\ehl). Our non-detection of the flow does not set a firm
constraint on the density of the accretion flow, since at such low
luminosities even the relatively massive disc predicted by the ADIOS
model in \cite{2014ApJ...797...92C} will be very optically thin to
electron scattering ($\tau \sim 10^{-3}$). However, if the orbital
inclination is high enough and the scale height of the flow stays
roughly constant, the optical depth through the flow should increase
with the accretion rate, so that by observing nearby neutron stars in
a low state ($\sim$$10^{-4}$--$10^{-3}L_{\rm Edd}$) we may see the
emission from the accretion flow and partly obscured emission from the
star. It may thus be possible to detect power-law emission from both
the central star and the accretion flow. Since the gas in the flow and
the surface could have different temperatures and densities, this
could manifest in the hard X-rays as two power-law components: one
with a break in the hard X-rays from the spectral cut-off of the
surface emission (as observed here) and a second power-law component
with a lower intensity extending to higher energies (from the
moderately optically thin accretion flow). A firm detection of
multiple power-law components would put strong constraints on the
radiative efficiency of the flow, since it would constrain the
relative intensities of the components.

Thermal radiation from the surface of the star could also be used to
constrain the optical depth of the accretion flow. Assuming that the
entire surface of the star is radiating (which seems reasonable for
sources like \cen, which show no clear magnetic activity), a small
inferred radiating area could indicate that a considerable amount of
emission from the star is being electron scattered by the accretion
flow, thus setting a constraint on the effective optical depth and
(where the system's inclination is known) the scale height of the
flow. It might also be possible to use changes in the pulsation
fraction with changing accretion rate in accreting millisecond X-ray
pulsars to constrain the optical depth of the intervening flow.  In
these systems a decrease in pulsation fraction could imply an
increased optical depth of the accretion flow, and be used to
constrain the optical depth as a function of luminosity.

Since the present observation of \cen\ shows both a thermal and
power-law component from the surface of the star, good modelling is
vital to understand how accreting matter hits the neutron star surface
and releases its energy. As is clear from Section~\ref{sec:models},
the total lack of a boundary layer model for low-accretion-rate
neutron star systems means constraining even the most basic properties
of boundary layer is only possible in a very qualitative way. Since
RIAFs are generally expected to be geometrically thick, they
generically have less specific angular momentum than thin discs, so
that the flow could bombard the star on mostly radial orbits. Particle
bombardment of the neutron star surface was studied
by \cite{1995ApJ...439..849Z} and \cite{2001A&A...377..955D}, and it was
found that particle bombardment leads to efficient thermalization of
the accretion energy so that the surface emits energy as a modified
blackbody. The work of \cite{2001A&A...377..955D} additionally found a
power-law component of varying strength, radiating either by
bremsstrahlung or inverse Compton scattering.

In contrast, the work of \cite{1999AstL...25..269I} and
\cite{2001ApJ...547..355P} considered a disc accreting on to a
boundary layer (where the gas in the boundary layer must redistribute
large quantities of angular momentum in order to settle on to the
star) and found that most of the accretion energy is released as power-law
radiation from a hot corona around the star. If these models are
correct and valid even at the low accretion rates seen in \cen, then
the thermal/power-law radiation balance could be a result of the
angular momentum distribution from the accretion flow, and could thus
be used to put an independent constraint on its scale height. Of
course, there may also be radiative coupling between the hot corona
and the cool neutron star surface, whereby the accretion energy is
radiated by the corona, which then heats the neutron star's
surface. This would require a very low albedo for the neutron star's
surface. In a future paper we will investigate some of these questions
more thoroughly in light of the quiescent observations of \cen,
but more detailed theoretical modelling of the boundary layer
structure is clearly warranted.

Understanding how the thermal component is produced in \cen\ is also
important for observations of neutron star cooling, which track the
thermal emission from quiescent neutron stars after an extended
outburst and fit them with crustal cooling models. Some of these
sources also show a power-law component in quiescence. As we argue in
this paper, the power-law component in all quiescent neutron stars is
most likely generated very close to the surface of the star, and this
accretion on to the surface could result in enhanced quasi-blackbody
emission. This extra source of energy may complicate the
interpretation of crustal cooling for quiescent neutron stars, since
this interpretation assumes that the surface temperature of the
neutron star is determined only by the cooling of the crust. \bhl The
possibility of continued accretion on the cooling neutron star has
been briefly discussed by e.g.\ \cite{2004ApJ...606L..61W},
\cite{2006MNRAS.372..479C}, and \cite{2011ApJ...736..162F}. A recent
study of how the spectra of neutron star LMXBs evolve with decreasing
luminosity has also suggested that quiescent neutron star spectra
(both the quasi-blackbody and power-law components) are dominated by
surface emission \citep{2014arXiv1409.6265W}.\ehl

Does \cen\ have a substantial dipolar magnetic field? For a spin
period of several milliseconds (like accreting millisecond X-ray
pulsars), assuming equation (\ref{eq:rm}) still applies, even a very
weak field ($\sim$$10^6$
G) will truncate the disc for $\dot{M}
\sim 10^{-6} \dot{M}_{\rm
  Edd}$ and channel the flow on to the stellar surface. This should
most likely produce some periodic modulation, but none is detected
down to a level of $\sim$$6.4$
per cent. The `strong propeller' scenario proposed by
\cite{2001ApJ...557..304M} required a field of at least $10^9$~G,
which is about an order of magnitude larger than typically seen in
accreting millisecond X-ray pulsars. While it is possible that the
magnetic geometry of \cen\ is such that pulsations are not seen, the
relatively large number of quiescent neutron stars without pulsations
suggest that this is unlikely to be the case. If LMXBs really do not
have strong dipolar fields, it implies a range of at least nine orders
of magnitude in the magnetic field strengths of neutron stars.

\section{\hspace{-0.1cm}\an{Conclusions}}
\label{sec:conclusion}
In this paper we have directly demonstrated that the accretion flow
around the quiescent neutron star binary \cen\ is most likely
undetected and therefore intrinsically radiatively inefficient. We
have argued that the near-balance between thermal and power-law
components in the X-ray spectrum, the relatively rapid covariation of
these components, and the shape of the power-law component
(specifically, its hardness and cut-off around 10~keV) all suggest
that the entire X-ray spectrum is generated in a boundary layer close
to the star. The accretion flow itself is therefore undetected, down
to a level of $\sim$6--34 per cent of the total observed luminosity,
directly demonstrating that the flow is most likely radiatively
inefficient (within uncertainties of the neutron star properties), as
has been postulated by various theoretical work. We have demonstrated
how this can set some (modest) constraints on the radiative
inefficiency of the accretion flow, and suggested ways in which
observations at somewhat higher luminosities with \nust\ could be used
to look for two power-law components (with different cut-off energies)
in the hard X-ray radiation, which would more clearly constrain the radiative
efficiency of the flow.

We conclude that there is no evidence of a substantial magnetic field
in \cen, based on a lack of detection of pulsations down to 6.4 per
cent of the total flux, and therefore agree with the conclusion of
\cite{2013MNRAS.436.2465B} that a `strong magnetocentrifugal
propeller' (as proposed by \citealt{2001ApJ...557..304M}) is
disfavoured. This would imply a span of at least nine orders of
magnitude in the dipolar strength of neutron stars. In the absence of
a magnetic propeller, and given the relatively large rate of mass
transfer from the companion star, the low luminosity of \cen\ strongly
favours RIAF models in which the majority of the accreting gas is
prevented from reaching the inner regions near the star, and could be
expelled in an outflow (ADIOS), recycled outward while remaining
gravitationally bound (CDAF), or inhibited by a magnetic field
(`magnetically arrested'). We propose using the luminosity versus
measured temperature of the blackbody surface component at somewhat
higher luminosities to constrain the optical depth of the accretion
flow.

We used simple physical arguments for the power-law component to
constrain the density of radiating gas: $10^{18} \lesssim n_{\rm e} \lesssim 10^{23}
{\rm~cm^{-3}}$. The measured electron temperature and predicted
time-scales for heating, cooling, and electron--proton scattering suggest
that the plasma likely has a single temperature.

Our conclusion is based on an empirical analysis of the data; a better
theoretical understanding of how the accretion energy is released in
the boundary layer (and in particular how the thermal component is
generated) is necessary to properly understand the implications of our
results on how observations of cooling neutron stars should be
interpreted, as well as to possibly set additional constraints on the
properties of the accreting matter (e.g. its temperature and angular
momentum) to further constrain the nature of the RIAF. A more thorough
investigation of how accretion flows interact with the surface layers
of neutron stars is therefore warranted.

\section*{\an{Acknowledgements}}
CRD'A and AP are financially supported by an NWO Vidi grant (PI:
Patruno). We are grateful to Daniel Wik for sharing his
\textsc{nuskybgd} background analysis tools for \nust. AP acknowledges
very useful conversations with Anne Archibald. CRD'A acknowledges very
useful conversations with Deepto Chakrabarty, Marat Gilfanov, Craig
Heinke, Andrew King, John Raymond, and Rudy Wijnands.

\bibliographystyle{mn2e}
\setlength\bibhang{1.4pc}
\renewcommand{\bibfont}{\footnotesize}
\balance

\bibliography{magbib_jkf}

\bsp

\label{lastpage}

\end{document}